\RequirePackage{fix-cm}

\documentclass[twocolumn,epjc3]{svjour3}
\usepackage{amsmath,amssymb}
\usepackage{epsfig,graphicx,widetext}
\smartqed  
\RequirePackage{graphicx}
\journalname{Eur. Phys. J. C}
\begin{document}

\title{Origin of inflation in CFT driven cosmology: $R^2$-gravity and non-minimally coupled inflaton models}

\titlerunning{Origin of inflation in CFT driven cosmology}        

\author{A. O. Barvinsky\thanksref{e1,addr1,addr2,addr3},
A. Yu. Kamenshchik\thanksref{e2,addr4,addr5} 
\and
D. V. Nesterov\thanksref{e3,addr1}}

\thankstext{e1}{e-mail: barvin@td.lpi.ru}
\thankstext{e2}{e-mail: kamenshchik@bo.infn.it}
\thankstext{e3}{e-mail: nesterov@td.lpi.it}

\institute{Theory Department, Lebedev
Physics Institute,
Leninsky Prospect 53, Moscow 119991, Russia \label{addr1}
\and
Tomsk State University, Department of Physics, Lenin Ave. 36, Tomsk 634050, Russia\label{addr2}
\and
Pacific Institute for Theoretical Physics, Department of Physics and Astronomy, UBC, 6224 Agricultural Road, Vancouver, BC V6T1Z1, Canada
\label{addr3}
\and
Dipartimento di Fisica e Astronomia, Universit\`a di Bologna and INFN,
via Irnerio 46, 40126 Bologna, Italy \label{addr4}
\and
L. D. Landau Institute for
Theoretical Physics, Moscow 119334, Russia\label{addr5}}
\date{Received: date / Accepted: date}

\maketitle
\begin{abstract}
We present a detailed derivation of the recently suggested new type of hill-top inflation originating from the microcanonical density matrix initial conditions in cosmology driven by conformal field theory (CFT)  [arXiv:1509.07270]. The cosmological instantons of topology  $S^1\times S^3$, which set up these initial conditions, have the shape of a garland with multiple periodic oscillations of the scale factor of the spatial $S^3$-section. They describe underbarrier oscillations of the inflaton and scale factor in the vicinity of the inflaton potential maximum, which gives a sufficient amount of inflation required by the known CMB data. We build the approximation of two coupled harmonic oscillators for these garland instantons and show that they can generate inflation consistent with the parameters of the CMB primordial power spectrum in the non-minimal Higgs inflation model and in $R^2$ gravity. In particular, the instanton solutions provide smallness of inflationary slow-roll parameters $\epsilon$ and $\eta<0$ and their relation $\epsilon\sim\eta^2$ characteristic of these two models. We present the mechanism of formation of hill-like inflaton potentials, which is based on logarithmic loop corrections to the asymptotically shift-invariant tree level potentials of these models in the Einstein frame. We also discuss the role of $R^2$-gravity as an indispensable finite renormalization tool in the CFT driven cosmology, which guarantees the non-dynamical (ghost free) nature of its scale factor and special properties of its cosmological garland type instantons. Finally, as a solution to the problem of hierarchy between the Planckian scale and the inflation scale we discuss the concept of a hidden sector of conformal higher spin fields.
\end{abstract}

\section{Introduction}
The problem of initial conditions in cosmology \cite{noboundary,tunnel} as a source of the inflationary scenario starts attracting attention again. A considerable success in explaining the CMB data \cite{WMAP,Planck} by amplification of primordial quantum cosmological pertubations \cite{ChibisovMukhanov} in the model of $R^2$ gravity \cite{Starobinskymodel} and the non-minimal Higgs inflation model \cite{BezShap,we} is called in question by objections against the origin of inflation \cite{Steinhardt}. These objections are based on the statement that distributions of initial conditions violate a widely accepted naturalness assumption that all forms of inflaton energy (kinetic, gradient and potential) should initially have the same Planckian scale magnitude \cite{Linde,Gorbunov}. Three known sources of these distributions -- pure no-boundary \cite{noboundary} and ``tunneling" \cite{tunnel} quantum states of the Universe and the Fokker-Planck equation for coarse-grained cosmological evolution \cite{Fokker-Planck} -- in their turn suffer from intrinsic difficulties associated with a missing clear canonical quantization ground, insufficient amount of generated inflation, anthropic (observer dependence) problems \cite{replica} and rather contrived multiverse measure problem \cite{sad}, etc.

The goal of this work is to give a detailed derivation of a recently suggested new model of hill-top inflation \cite{hill-top} in the  theory of CFT driven cosmology \cite{slih,why} which is likely to circumvent the difficulties of the above type. This theory represents the synthesis of two main ideas -- new concept of the cosmological microcanonical density matrix as the initial state of the Universe and application of this concept to the system with a large number of quantum fields conformally coupled to gravity. It plays an important role within the cosmological constant and dark energy problems. In particular, its statistical ensemble is bounded to a finite range of values of the effective cosmological constant, it incorporates inflationary stage and is potentially capable of generating the cosmological acceleration phenomenon within the so-called Big Boost scenario \cite{slih,bigboost}. Moreover, as was noticed in \cite{DGP/CFT}, the CFT driven cosmology provides perhaps the first example of the initial quantum state of the inflationary Universe, which has a thermal nature of the primordial power spectrum of cosmological perturbations \cite{CMBA-theorem}. This suggests a new mechanism for the red tilt of the CMB anisotropy, complementary to the conventional mechanism which is based on a small deviation of the inflationary expansion from the exact de Sitter evolution \cite{ChibisovMukhanov}.

This setup has a clear origin in terms of operator quantization of gravity theory in the Lorentzian signature spacetime and is based on a natural notion of the microcanonical density matrix as a projector on the space of solutions of the quantum gravitational Dirac constraints -- the system of the Wheeler-DeWitt equations \cite{why,whyBFV}. Its statistical sum has a representation of the Euclidean quantum gravity (EQG) path integral \cite{slih,why}
    \begin{eqnarray}
    &&Z=
    \!\!\int\limits_{\,\,\rm periodic}
    \!\!\!\! D[\,g_{\mu\nu},\varPhi\,]\;
    e^{-S[\,g_{\mu\nu},\varPhi\,]},         \label{Z}
    \end{eqnarray}
over metric $g_{\mu\nu}$ and matter fields $\varPhi$ which are
periodic on the Euclidean spacetime with a time compactified to a circle $S^1$.

As shown in \cite{slih,why}, this statistical sum is approximately calculable and has a good predictive power in the gravitational model with the primordial cosmological constant $\varLambda$ and the matter sector which mainly consists of a large number $\mathbb{N}$ of free (linear) fields $\phi$ conformally coupled to gravity -- conformal field theory (CFT) with the action $S_{CFT}[\,g_{\mu\nu},\varPhi\,]$,
    \begin{eqnarray}
    &&S[\,g_{\mu\nu},\varPhi\,]=-\frac{M_P^2}2
    \int d^4x\,g^{1/2}\,(R-2\varLambda)\nonumber \\
    &&\ \ \ \ \ \ \ \ \ \ +S_{CFT}[\,g_{\mu\nu},\varPhi\,].    \label{tree0}
    \end{eqnarray}
Important point, which allows one to overstep the limits of the usual semiclassical expansion, consists here in the possibility to omit the integration over conformally non-invariant matter fields and spatially-inhomogeneous metric modes on top of a dominant contribution of numerous conformal species. Integrating them out one obtains the effective gravitational action $S_{\rm eff}[\,g_{\mu\nu}]$ which differs from (\ref{tree0}) by $S_{CFT}[\,g_{\mu\nu},\varPhi\,]$ replaced with $\varGamma_{CFT}[\,g_{\mu\nu}]$ -- the effective action of $\varPhi$ on the background of $g_{\mu\nu}$,
    \begin{eqnarray}
    e^{-\varGamma_{C\!F\!T}[\,g_{\mu\nu}]}=\int D\varPhi\,
    e^{-S_{C\!F\!T}[\,g_{\mu\nu},\varPhi\,]}, \label{GammaCFT}
    \end{eqnarray}
On Friedmann-Robertson-Walker (FRW) background this action is exactly calculable by using the local conformal transformation to the static Einstein universe and well-known local trace anomaly. The resulting \\ $\varGamma_{CFT}[\,g_{\mu\nu}]$ turns out to be the sum of the anomaly contribution and free energy of conformal matter fields on the sphere $S^3$ at the temperature determined by the circumference of the compactified time dimension $S^1$.
\footnote{Note that in previous works on trace anomaly applications in cosmology the contribution of the static Einstein universe was either restricted to the case of the vacuum state (Casimir energy) \cite{Casimir_energy} or recovered for a particular value of integration constant in the stress tensor conservation law \cite{FHH,previous}. Here it follows directly from the density matrix prescription for the initial state of the Universe, which endows the Einstein universe with the radiation gas of CFT particles and makes our setting of the problem essentially different from previous studies.}
This is the main calculational advantage provided by the local Weyl invariance of $\varPhi$ conformally coupled to $g_{\mu\nu}$.

Physics of the CFT driven cosmology is entirely determined by this effective action. Solutions of its equations of motion, which give a dominant contribution to the statistical sum, are the cosmological instantons of $S^1\times S^3$ topology, which have the Friedmann-Robertson-Walker metric
    \begin{eqnarray}
    g_{\mu\nu}^{FRW}dx^\mu dx^\nu=N^2(\tau)\,d\tau^2
    +a^2(\tau)\,d^2\Omega^{(3)}              \label{FRW}
    \end{eqnarray}
with a periodic lapse function $N(\tau)$ and scale factor $a(\tau)$ -- functions of the Euclidean time belonging to the circle $S^1$ \cite{slih}. These instantons serve as initial conditions for the cosmological evolution $a_L(t)$ in the physical Lorentzian spacetime. The latter follows from $a(\tau)$ by analytic continuation $a_L(t)=a(\tau_*+it)$ at the point of the maximum value of the Euclidean scale factor $a_+=a(\tau_*)$. The fact that these instantons exist only in the finite range of $\varLambda$ implies the restriction of the microcanonical ensemble of universes to this range, which from the viewpoint of string theory can be interpreted as the solution of the landscape problem for stringy vacua.

As was originally mentioned in \cite{bigboost} this scenario can incorporate a finite inflationary stage if the model (\ref{tree0}) is generalized to the case when $\varLambda$ is replaced by a composite operator $\varLambda(\phi)=V(\phi)/M_P^2$ -- the potential of the inflaton $\phi$ slowly varying during the Euclidean and inflationary stages and decaying in the end of inflation by a usual exit scenario. The goal of this paper is to develop such a generalization which starts with the replacement of (\ref{tree0}), $S[\,g_{\mu\nu},\varPhi\,] \to S[\,g_{\mu\nu},\phi,\varPhi\,]$, by the action with the inflaton field,
    \begin{eqnarray}
    &&S[\,g_{\mu\nu},\phi,\varPhi\,]=
    \int d^4x\,g^{1/2}\,\left(-\frac{M_P^2}2\,R\right.\nonumber\\
    &&\qquad\qquad\left.+\frac12(\nabla\phi)^2+V(\phi)\right)
    +S_{CFT}[\,g_{\mu\nu},\varPhi\,],         \label{tree1}
    \end{eqnarray}
whose potential $V(\phi)$ simulates the effect of the primordial cosmological constant. Under this replacement the restriction of the primordial cosmological constant range of the above type, $\varLambda(\phi)=V(\phi)/M_P^2$, becomes a selection of the range of $\phi$ or fixation of the initial conditions for inflation. These initial conditions is a principal goal of this paper.

As we will see CFT driven cosmology realizes these initial conditions in the form of the new type of hill-top inflation originating from the underbarrier oscillations of the inflaton $\phi$ and the scale factor $a$ in the vicinity of local maxima of $V(\phi)$. Thus, these initial conditions are not the Planckian scale parameters introduced by hands from some ad hoc arguments of naturalness \cite{Linde,Steinhardt}, but rather become derivable from microcanonical partition function as calculable parameters of the corresponding saddle point configurations.

Dynamical nature of the physical mode simulating the effect of the cosmological constant has also another important implication. The properties
of instantons in the CFT driven cosmology critically depend on the vanishing of one of the conformal anomaly parameters (the coefficient of $\Box R$). It is renormalization ambiguous, because it can be arbitrarily changed by the
addition of the local $R^2$ term, and therefore can be renormalized to zero to provide needed properties of CFT initial conditions \cite{slih,hill-top}.
Quite interestingly, this finite renormalization goes beyond an ad hoc
assumption of \cite{slih} and can be a result of inclusion of the Starobinsky $R^2$ model, one part of which would provide this renormalization in the anomaly part of the action and the other part would give a scalar mode simulating the effect of the inflaton field -- effective cosmological term decaying at the exit from inflation.

The plan of the paper is as follows. In Sect.2 we overview the model of CFT driven cosmology with a fundamental cosmological constant. Sect.3 is devoted to the generalization of this model to the case of the dynamical inflaton field minimally coupled to gravity. It suggests the approximation of coupled harmonic oscillators for cosmological instantons in the regime of the Euclidean ``slow roll", which set up initial conditions for inflation in physical spacetime, starting in the vicinity of the top of the inflaton potential. Sect.4 demonstrates formation of the relevant hill-like effective potential in the model of Higgs inflation with a strong non-minimal coupling of the inflaton to the scalar curvature. In Sect. 5 a similar mechanism is discussed for the Starobinsky model of $R^2$-gravity which is shown to play a double role in the CFT scenario: finite renormalization of the quantum effective action, which provides special properties of the cosmological instantons, and generation of the dynamical inflaton feeding the CFT scenario with the hill-like potential. Sect.6 contains an attempt to solve the hierarchy problem in the CFT cosmology by using a hidden sector of numerous conformal higher spin fields and the concluding Sect.7 briefly discusses observational prospects of the model. Two appendices contain technical details of the harmonic oscillator approximation and its Euclidean ``slow-roll" regime.

\section{Model with a fundamental cosmological constant}

For cosmology with $S^1\times S^3$ topology and FRW metric (\ref{FRW}) its effective action $S_{\rm eff}[\,g_{\mu\nu}^{FRW}]\equiv S_{\rm eff}[\,a,N\,]$ reads \cite{slih}
    \begin{eqnarray}
    &&S_{\rm eff}[\,a,N\,]=6\pi^2 M_P^2\int_{S^1} d\tau\,N \left\{-aa'^2
    -a+ \frac\varLambda3 a^3\right.\nonumber \\
    &&\left.+\,B\left(\frac{a'^2}{a}
    -\frac{a'^4}{6 a}\right)
    +\frac{B}{2a}\,\right\}
    +F(\eta),                           \label{effaction0}\\
    &&F(\eta)=\pm\sum_{\omega}\ln\big(1\mp
    e^{-\omega\eta}\big),                 \label{freeenergy}\\
    &&\eta=\int_{S^1} \frac{d\tau N}a,     \label{period}
    \end{eqnarray}
where $a'\equiv da/Nd\tau$. The first three terms in curly brackets of (\ref{effaction0}) represent the Einstein action with a fundamental cosmological constant $\varLambda\equiv 3H^2$ ($H$ is the corresponding Hubble parameter). The constant $B$ is a coefficient of the contributions of the conformal anomaly and vacuum (Casimir) energy $(B/2a)$ on a conformally related static Einstein spacetime mentioned in Introduction. This constant,
    \begin{eqnarray}
    B=\frac{\beta}{8\pi^2 M_P^2},         \label{B}
    \end{eqnarray}
expresses via the coefficient $\beta$ of the Gauss-Bonnet term $E=R_{\mu\nu\alpha\gamma}^2-4R_{\mu\nu}^2+ R^2$ in the trace anomaly of conformal matter fields
    \begin{eqnarray}
    &&g_{\mu\nu}\frac{\delta
    \varGamma}{\delta g_{\mu\nu}} =
    \frac{1}{4(4\pi)^2}g^{1/2}
    \left(\alpha \Box R +
    \beta E +
    \gamma C_{\mu\nu\alpha\beta}^2\right). \label{anomaly}
    \end{eqnarray}
It should be emphasised here that this effective action is independent of the anomaly coefficients $\alpha$ and $\gamma$, because it is assumed that $\alpha$ is renormalized  to zero by a local counterterm,
\begin{equation}
\varGamma_{CFT}\to\varGamma_{CFT}^R\equiv\varGamma_{CFT}
+\frac\alpha{384\pi^2}\int d^4x\,g^{1/2}R^2.
\label{renormaction}
\end{equation}
This guarantees absence of higher derivative terms in (\ref{effaction0}) -- non-ghost nature of the scale factor -- and simultaneously gives the renormalized Casimir energy a partiular value proportional to $B/2=\beta/16\pi^2M_P^2$ \cite{universality}. Both of these properties are critically important for the instanton solutions of effective equations. The coefficient $\gamma$ of the Weyl tensor term $C^2_{\mu\nu\alpha\beta}$ does not enter (\ref{effaction0}) because $C_{\mu\nu\alpha\beta}$ identically vanishes for any FRW metric.

Finally, $F(\eta)$ is the free energy of conformal fields also coming from this Einstein space -- a typical boson or fermion sum over CFT field oscillators with energies $\omega$ on a unit 3-sphere, $\eta$ playing the role of the inverse temperature --- an overall circumference of $S^1$ in the $S^1\times S^3$ instanton, calculated in units of the conformal time (\ref{period}).

The statistical sum (\ref{Z}) is dominated by the solutions of the effective equation, $\delta S_{\rm eff}/\delta N(\tau)=0$, which in the gauge $N=1$ reads
    \begin{eqnarray}
    &&-\frac{\dot a^2}{a^2}+\frac{1}{a^2}
    -B \left(\,\frac{\dot a^4}{2a^4}
    -\frac{\dot a^2}{a^4}\right) =
    \frac\varLambda3+\frac{C}{ a^4},\ \dot a=\frac{da}{d\tau},               \label{efeq0}\\
    &&C =
    \frac{B}2+\frac1{6\pi^2 M_P^2}\,\frac{dF}{d\eta}, \label{bootstrap}\\
    &&\frac{dF}{d\eta}=
    \sum_\omega\frac{\omega}
    {e^{\omega\eta}\mp 1}.
    \end{eqnarray}
This is the modification of the Euclidean Friedmann equation by the anomalous $B$-term and the radiation term $C/a^4$. The constant $C$ here characterizes the sum of the Casimir energy and the energy of thermally excited particles with the inverse temperature $\eta$ given by (\ref{period}).

This quadratic equation (\ref{efeq0}) can be solved for $\dot a^2$,
    \begin{eqnarray}
    \dot{a}^2 &=& \sqrt{\frac{(a^2-B)^2}{B^2}
    +\frac{2H^2}{B}\,(a_+^2-a^2)(a^2-a_-^2)}\nonumber \\
    &&-\frac{a^2-B}{B},                      \label{mainEq}\\
    a_\pm^2&\equiv&
    \frac{1\pm\sqrt{1-4CH^2}}{2H^2},          \label{apm}
    \end{eqnarray}
to give a periodic oscillation of $a$ between its maximal and minimal values $a_\pm$, provided that at $a_-$ we have a turning point with a vanishing $\dot a$, which means that $a_-^2>B$. This inequality immediately yields the first two restrictions on the range of $H^2$ and $C$
    \begin{eqnarray}
    &&H^2\leq\frac1{2B},      \label{Hbound}\\
    &&C\geq B-B^2H^2,     \label{lowerCbound}
    \end{eqnarray}
whereas the third one follows from the requirement of reality of turning points $a_\pm$,
    \begin{eqnarray}
    C\leq \frac1{4H^2}.    \label{upperCbound}
    \end{eqnarray}

As shown in \cite{slih,why,DGP/CFT} the solutions of this integro-differential equation\footnote{Note that the constant $C$ is a nonlocal functional of the history $a(\tau)$ -- Eq.(\ref{bootstrap}) plays the role of the bootstrap equation for the amount of radiation determined by the background on top of which this radiation evolves and produces back reaction.} give rise to the set of periodic $S^3\times S^1$ instantons with the oscillating scale factor -- {\em garlands} that can be regarded as the thermal version of the Hartle-Hawking instantons. The scale factor oscillates $m$ times ($m=1,2,3,...$) between the maximum and minimum values (\ref{apm}), $a_-\leq a(\tau)\leq a_+$, so that the full period of the conformal time (\ref{period}) is the $2m$-multiple of the integral between the two neighboring turning points of $a(\tau)$, $\dot a(\tau_\pm)=0$,
    \begin{eqnarray}
    &&\eta=2m\int_{a_-}^{a_+}
    \frac{da}{a'a}.                       \label{period1}
    \end{eqnarray}
This value of $\eta$ is finite and determines effective temperature $T=1/\eta$ as a function of $G=1/8\pi M_P^2$ and $\varLambda=3H^2$. This is the artifact of a microcanonical ensemble in cosmology \cite{why} with only two freely specifiable dimensional parameters --- the gravitational and cosmological constants.

\begin{figure}[h]
\centerline{\epsfxsize 9cm \epsfbox{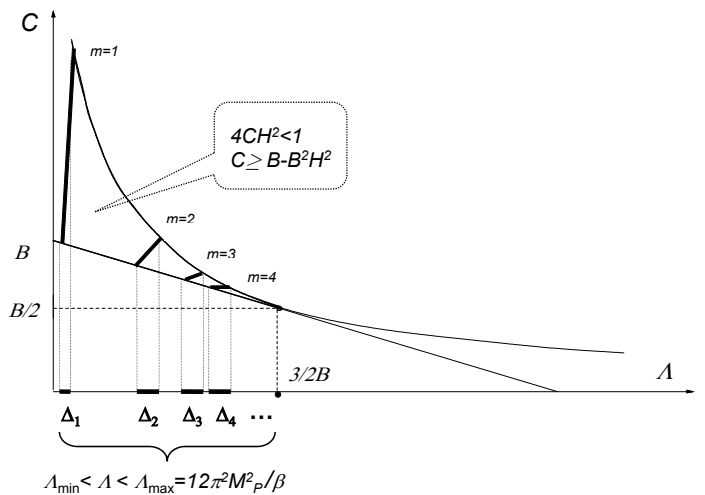}}
\caption{\small Band range of garland instantons formed by projections $\Delta_m$ of their $m$-folded one-parameter families onto the axis of $\varLambda=3H^2$.
 \label{Fig.1}}
\end{figure}

According to (\ref{Hbound}) these garland-type instantons exist only in the limited range of the cosmological constant $\varLambda=3H^2$ \cite{slih}. In view of (\ref{lowerCbound}) and (\ref{upperCbound}) they belong to the curvilinear domain in the two-dimensional plane of the Hubble constant $H^2$ and the amount of radiation constant $C$ (each instanton being represented by a point in this plane). In this domain they form a countable, $m=1,2,...$, sequence of one-parameter families -- curves interpolating between the lower straight line boundary $C=B-B^2H^2$ and the upper hyperbolic boundary $C=1/4H^2$. Each curve corresponds to a respective $m$-folded instantons of the above type. Therefore, the spectrum of admissible values of $\varLambda$ has a band structure, each band being a projection of the $m$'th curve to the $H^2$ axis. The sequence of bands of ever narrowing widths with $m\to\infty$ accumulates at the upper bound of this range $H^2_{\rm max}=1/2B$. The lower bound $H^2_{\rm min}$ -- the lowest point of $m=1$ family -- can be obtained numerically for any field content of the model.

For a large number of conformal fields $\mathbb{N}$, and therefore a large $\beta\propto \mathbb{N}$, the both bounds are of the order $m_P^2/\mathbb{N}$. Thus the restriction (\ref{Hbound}) suggests a kind of $1/\mathbb{N}$ solution of the cosmological constant problem, because specifying a sufficiently high number of conformal fields one can achieve a primordial value of $\varLambda$ well below the Planck scale where the effective theory applies, but high enough to generate a sufficiently long inflationary stage.

\section{Minimally coupled scalar field inflaton in CFT cosmology}

Generalization of the action (\ref{tree0}) to the effective $\varLambda$ generated by a scalar field implies the transition from the effective minisuperspace action (\ref{effaction0}) to
    \begin{eqnarray}
    &&\varGamma[\,a,N,\phi\,]=6\pi^2 M_P^2\oint d\tau\,N\,\left\{
    \vphantom{\frac11}-aa'^2
    -a\right.\nonumber \\
    &&\quad+B\left(\frac{a'^2}{a}
    -\frac{a'^4}{6 a}+\frac1{2a}\right)
    \left.+\frac{a^3}{3M_P^2}\,\left(V(\phi)+\frac12\,
    \phi'^2\right)\,\right\}\nonumber\\
    &&\quad+ F(\eta),\quad 
    a'=\frac1N\frac{da}{d\tau},\,\phi'
    =\frac1N\frac{d\phi}{d\tau},     \label{minimal_inflaton}
    \end{eqnarray}
which leads to the replacement
    \begin{eqnarray}
    H^2\to H^2
    = \frac1{3M_P^2}\left(V(\phi)
    -\frac{\dot\phi^2}2\right)        \label{replacement}
    \end{eqnarray}
in the effective Friedmann equation (\ref{efeq0}) and to the origin of dynamical equation for the inflaton $\phi$,
    \begin{eqnarray}
    &&-\frac{\dot a^2}{a^2}+\frac{1}{a^2}
    -B \left(\,\frac{\dot a^4}{2a^4}
    -\frac{\dot a^2}{a^4}\right)\nonumber \\
    &&\qquad\qquad\qquad=\frac1{3M_P^2}\left(V
    -\frac12\dot\phi^2\!\right)
    +\frac{C}{ a^4},                     \label{efeq}\\
    &&\frac1{a^3}\frac{d}{d\tau}\,
    \big(\,a^3\dot\phi\,\big)
    -\frac{\partial V}{\partial\phi}=0.    \label{infleq}
    \end{eqnarray}

The possibility of existence of periodic solutions of this system of equations qualitatively follows from counting its integration constants. Because of the fact that Eq.(\ref{efeq}) is of the first order in derivatives, their number is three -- the initial value of the inflaton $\phi_0$ at some $\tau=\tau_0$, its initial time derivative $\dot\phi_0$ and the initial vale of the scale factor $a_0$. Thus the solution has the form
    \begin{eqnarray}
    \phi(\tau)=\varPhi(\tau\,|\,\phi_0,\dot\phi_0,a_0),\quad
    a(\tau)=A(\tau\,|\,\phi_0,\dot\phi_0,a_0).
    \end{eqnarray}
Requirement of periodicity with some period $T$ implies the following equations
    \begin{eqnarray}
    &&\varPhi(\tau_0+T\,|\,\phi_0,\dot\phi_0,a_0)=\phi_0,\nonumber \\
    &&\frac{d}{dT}\varPhi(\tau_0+T\,|\,\phi_0,\dot\phi_0,a_0)
    =\dot\phi_0,\nonumber \\
    &&A(\tau_0+T\,|\,\phi_0,\dot\phi_0,a_0)=a_0.
    \end{eqnarray}
From these three equations one can determine three initial condition parameters as functions of the period,
    \begin{eqnarray}
    \phi_0,\dot\phi_0,a_0=\phi_0(T),\dot\phi_0(T),a_0(T),
    \end{eqnarray}
which are locally unique provided the nonvanishing Jacobian $\partial(\varPhi,\dot\varPhi,A)/\partial(\phi_0,\dot\phi_0,a_0)\neq 0$.

The period $T$ does not remain a free parameter, because of the additional requirement. Analytical continuation of the Euclidean solution to the {\em real} solution in the Lorentzian time $\tau=\tau_*+it$ requires that time derivatives of $\phi(\tau)$ and $a(\tau)$ at $\tau_*$ vanish. The existence of such $\tau_*$ is guaranteed for one of the variables, say $\Phi(\tau\,|\,\phi_0(T),\dot\phi_0(T),a_0(T)\,)$, $\dot\Phi(\tau_*\,|\,\phi_0(T),\dot\phi_0(T),a_0(T)\,)=0$, in view of its periodic motion away from $\phi_0$ and back to $\phi_0$ during the period of oscillations. Now if we demand that at the same moment the time derivative of $a(\tau)$ is also zero, then we get the equation for $T$,
    \begin{eqnarray}
    \dot A(\tau_*\,|\,\phi_0(T),\dot\phi_0(T),a_0(T)\,)=0,
    \end{eqnarray}
which gives a locally unique solution $T=T(\tau_*)$. This counting of constants of motion guarantees that, if some approximate periodic solution satisfying the condition of Euclidean-Lorentzian junction exists, then perturbation theory will guarantee the periodicity of perturbed solution. Let us construct such an approximate solution which will serve as a lowest order of perturbation theory for the exact periodic trajectory.

To begin with, note that periodicity imposes a restriction on the shape of the potential $V(\phi)$ -- integration of Eq.(\ref{infleq}) over the period gives
    \begin{eqnarray}
    \oint d\tau\,a^3\frac{\partial V}{\partial\phi}=0.
    \end{eqnarray}
This implies that the gradient of $V$ changes sign at some point $\phi_0$,
    \begin{eqnarray}
    \frac{\partial V(\phi_0)}{\partial\phi_0}\equiv V'_0=0,
    \end{eqnarray}
so that inflaton oscillations occur in the vicinity of the extremum of $V$. Critical point is that these oscillations can take place only in the vicinity of the potential {\em maximum} in the underbarrier regime, which is shown schematically in Fig.2. This follows from a simple fact that our equations of motion are in Euclidean time, and they admit periodic oscillations of the scale factor only in the Euclidean time. Therefore, inflaton oscillations take place also in the underbarrier regime, which implies the vicinity of a potential maximum.

\begin{figure}[h]
\centerline{\epsfxsize 9cm \epsfbox{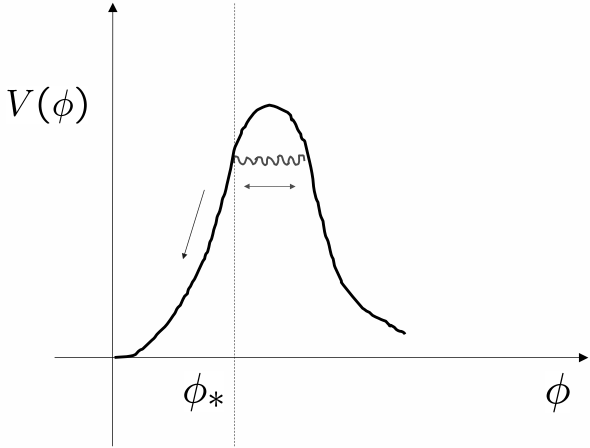}}
\caption{\small Picture of hill-top inflation: underbarrier oscillations indicated by waverly line give rise to inflationary slow roll at the turning point $\phi_*$.
 \label{Fig.2}}
\end{figure}

Critical difference from the tunneling prescription of \cite{noboundary,tunnel} here is that selection of the inflaton potential maxima follows not from the properties of the probability distribution, but from the requirement of a periodic solution for a saddle point of the partition function path integral. This solution gives rise to a new type of hill-top inflation which starts at the turning point $\phi_*$ on the slope of the potential close to its maximum.

\subsection{Harmonic oscillator approximation}

It is obvious that solutions described above for a constant primordial $\varLambda$ remain approximately true for the system of equations (\ref{efeq})-(\ref{infleq}) when the inflaton field and its effective cosmological ``constant" (\ref{replacement}) do not vary much and their variations are not rapid. This is consistent with sufficiently slow variations of the scale factor, because from the inflaton equation of motion (\ref{infleq}) it follows that the relevant rate of change of (\ref{replacement}) is also small for small $\dot\phi$ and $\dot a$,
    \begin{eqnarray}
    \frac{d}{d\tau}
    \left(V-\frac{\dot\phi^2}2\right)=
    3\frac{\dot a}a\,\dot\phi^2.             \label{dotH}
    \end{eqnarray}

If we disregard the friction term in (\ref{infleq}) and
expand the potential in the vicinity of its maximum at $\phi_0$,
    \begin{eqnarray}
    &&V(\phi)\simeq V_0-\frac12\,\mu^2
    (\phi-\phi_0)^2,\quad V_0=
    V(\phi_0),\nonumber\\
    &&\mu^2=
    -\frac{d^2V(\phi_0)}{d\phi_0^2}  \label{quadpot}
    \equiv-V''_0>0,
    \end{eqnarray}
small oscillations of $\phi$ will be nearly harmonic,
    \begin{eqnarray}
    &&\ddot\phi+\mu^2(\phi-\phi_0)\simeq 0,\\
    &&\phi=\phi_0-\Delta_\phi\cos(\mu\tau),  \label{harmonicphi}
    \end{eqnarray}
with some amplitude $\Delta_\phi=\phi_0-\phi_*$, where $\phi_*$ is a turning point shown on Fig.2.

A related quasi-harmonic behaviour of the scale factor is analytically available for solutions close to the upper hyperbolic boundary of the domain (\ref{upperCbound}) of Fig.1. In a small vicinity of this boundary smallness of $\dot a$ is guaranteed, because at $C=1/4H^2$ the scale factor is pinched between the coincident $a_-$ and $a_+$, $a(\tau)=a_-=a_+$. If we use the notations introduced in \cite{one-loop}
    \begin{eqnarray}
    &&\varepsilon=1-2BH^2,\quad
    0\leq\varepsilon\leq 1,     \label{varepsilon}\\
    &&\Delta\equiv d\varepsilon=\sqrt{1-4CH^2},
    \quad 0\leq \Delta\leq \varepsilon,       \label{Delta}
    \end{eqnarray}
then the proximity to the upper boundary is determined by smallness of $\Delta$, $\Delta\ll 1$. In these variables the admissible domain of solutions looks as a unit triangle (the quadrangle in terms of the quantities $\varepsilon$ and $d\equiv\Delta/\varepsilon$ of \cite{one-loop}) and the scale factor oscillating between the turning points $a_\pm$ can be parameterized by the variable $y$ running between $-1$ and $1$ according to
    \begin{eqnarray}
    \frac{a^2}B=
    \frac{1+y\Delta}{1-\varepsilon}.   \label{a_vs_y}
    \end{eqnarray}
In terms of $y$ the equation (\ref{mainEq}) for $\dot a^2$ reads as
    \begin{eqnarray}
    &&\!\!\!\!\dot a^2=\frac{\Delta^2}\varepsilon\,(1-y^2)\nonumber\\
    &&\!\!\!\!\times\frac1
    {\big[\,(1+\frac{y\Delta}\varepsilon)^2
    +(\frac\Delta\varepsilon)^2(1-\varepsilon)(1-y^2)\,
    \big]^{1/2}+1+\frac{\Delta y}\varepsilon}.                 \label{dota}
    \end{eqnarray}
All these equations hold also for the inflaton model. In this case the parameters $\varepsilon$ and $\Delta$, depending on $H^2$, become the functions of time. However, if they are sufficiently slowly varying, time derivatives $\dot\varepsilon$ and $\dot\Delta$ can be disregarded. Then the above equation for $\dot a$ reads as the equation for $\dot y$

    \begin{eqnarray}
    &&\!\!\!\!\dot y^2=\frac{1-\varepsilon}{\varepsilon B}\,
    (1-y^2)\,\nonumber\\
    &&\!\!\!\!\times\frac{1+y\Delta}
    {\big[\,(1+\frac{y\Delta}\varepsilon)^2
    +(\frac\Delta\varepsilon)^2(1-\varepsilon)(1-y^2)\,
    \big]^{1/2}+1+\frac{\Delta y}\varepsilon}.       \label{doty}
    \end{eqnarray}
Therefore, in the domain of small $\Delta$
    \begin{eqnarray}
    \Delta\ll \varepsilon,                       \label{approximation0}
    \end{eqnarray}
where
    \begin{eqnarray}
    &&\!\!\!\!\frac{1+y\Delta}
    {\big[\,(1+y\Delta/\varepsilon)^2
    +(\Delta/\varepsilon)^2(1-\varepsilon)(1-y^2)\,
    \big]^{1/2}+1+y\Delta/\varepsilon}\nonumber\\
    &&\qquad\qquad\qquad
    =\frac12+O\left(\frac\Delta\varepsilon\right),  \label{approximation00}
    \end{eqnarray}
the variable $y$ describes a harmonic oscillator with a slowly varying frequency $\omega$ and a unit amplitude,
    \begin{eqnarray}
    &&\dot y^2+\omega^2y^2=\omega^2, \\
    &&\omega^2=2\frac{1-\varepsilon}
    {\varepsilon B}
    =\frac{4H^2}{1-2BH^2},            \label{omega}\\
    &&y=\cos(\omega\tau),         \label{y}
    \end{eqnarray}
so that
    \begin{eqnarray}
    a^2=\frac1{2H^2}
    +\frac{\Delta}{2H^2}\,\cos(\omega\tau). \label{harmonica}
    \end{eqnarray}

Conditions of applicability of this approximation, that should guarantee smallness of the friction term in the inflaton oscillator and a small change of $\varepsilon$ and $H$ during the period of oscillation $2\pi/\omega$, are
    \begin{eqnarray}
    &&|\,\ddot\phi\,|\gg \left|\,3\frac{\dot a}a\,\dot\phi\,\right|,\\
    &&\omega H^2\gg \left|\,\frac{d}{d\tau}H^2\,\right|=\frac1{3M_P^2} \left|\,\frac{d}{d\tau}
    \left(V-\frac{\dot\phi^2}2\right)\right|\nonumber \\
    &&\qquad\qquad\qquad\quad=
    \frac1{M_P^2}
    \left|\,\frac{\dot a}a\,\dot\phi^2\right|.
    \end{eqnarray}
Since $\dot a/a\sim \omega \Delta/2$, these bounds lead to
    \begin{eqnarray}
    &&\mu\gg \omega\Delta,         \label{bound1}\\
    &&W\Delta\ll 1,       \label{bound2}
    \end{eqnarray}
where we have introduced the notation
    \begin{eqnarray}
    W\equiv\frac{\dot\phi^2}{2M_P^2H^2}
    \sim \frac{\Delta_\phi^2}{M_P^2}
    \frac{\mu^2}{H^2},                      \label{W}
    \end{eqnarray}

In addition to the above bounds we also need a small rate of change of $\varepsilon$ and $\Delta$ necessary for the transition from Eq.(\ref{dota}) to Eq.(\ref{doty}). Smallness of $\dot\varepsilon$ is guaranteed by the bound (\ref{bound2}) derived above,
    \begin{eqnarray}
    \frac{\Delta_\phi^2}{M_P^2}
    \frac{\mu^2}{H^2}\ll \frac1\Delta,       \label{bound20}
    \end{eqnarray}
whereas a small $|\dot\Delta|\ll\omega\Delta$ requires a much stronger bound, because $\dot\Delta\simeq-(1/\Delta)W\dot a/a\sim W\omega\ll\omega\Delta$ or
    \begin{eqnarray}
    \frac{\Delta_\phi^2}{M_P^2}
    \frac{\mu^2}{H^2}\ll \Delta.       \label{bound3}
    \end{eqnarray}

Interestingly, the last bound -- smallness of $\dot\Delta$ -- can be replaced by the opposite limit $W\gg\Delta$. For large $\dot\Delta$ Eq.(\ref{doty}) does not hold, and oscillations of $y$ become strongly anharmonic. However, as shown in Appendix A, the period of these oscillations remains the same $2\pi/\omega$ and $y$ behaves as $y\simeq\left|\,\sin\frac{\omega\tau}2\right|$ provided the function (\ref{W}) is slowly varying compared to the oscillations of the scale factor, $\dot W\ll\omega W$. This leads to the extra bound on $\mu$, $\mu\ll\omega$, so that together with (\ref{bound1}) the frequency of inflaton oscillations $\mu$ belongs to a limited range
    \begin{eqnarray}
    \omega\Delta\ll\mu\ll\omega ,       \label{bound4}
    \end{eqnarray}
which is nonempty for assumed values of $\Delta\ll 1$.

Thus, the inflaton field and the scale factor represent two coupled quasi-harmonic oscillators with frequencies $\mu$ and $\omega$. Periodicity of their motion implies their frequencies to be commensurable
    \begin{eqnarray}
    m\mu=n\omega,              \label{commensurable}
    \end{eqnarray}
where $m$ and $n$ are some integer numbers. Therefore, from the definition (\ref{omega}) of the frequency $\omega$ it follows that
    \begin{eqnarray}
    H^2=\frac1{2B+4n^2/\mu^2m^2}.    \label{Hmn}
    \end{eqnarray}

Thus the approximation of two coupled oscillators works for a wide range of instantons lying close to the upper hyperbolic boundary even when the parameter $\Delta$ is rapidly varying in time -- as fast as the the scale factor and violating the bound (\ref{bound3}). The lower bound for $\mu$ in (\ref{bound4}) implies fast oscillations of $a$ with a slower motion of $\phi$, $m\gg n\geq 1$. This conclusion is also confirmed by considering the slow roll smallness parameters for the inflation stage originating from the transition through the turning point $\phi_*$.

\subsection{Inflation stage and its slow roll parameters}

Hill-top inflation histories in Lorentzian time, $\phi_L(t)$ and $a_L(t)$, originate by analytic continuation of the Euclidean solutions (\ref{harmonicphi}) and (\ref{harmonica}) to $\tau=2m\pi/\omega+it$, where $m\gg 1$ is the number of oscillations of the scale factor in the garland instanton. For a small time $t$ the linearized Lorentzian solutions read
    \begin{eqnarray}
    &&\phi_L(t)=\phi_0-\Delta_\phi\cosh(\mu t),\\ &&a^2_L(t)=\frac1{2H^2}+\frac{\Delta}{2H^2}\cosh(\omega t),\quad
    \omega t\ll 1,       \label{aL}
    \end{eqnarray}
while at later times nonlinear effects start dominating, so that the Lorentzian version of the nonlinear equation (\ref{efeq}) with  $\varLambda$ replaced by the dynamical inflaton energy density,
    \begin{eqnarray}
    \varLambda\to\frac1{M_P^2}\Big(V(\phi_L)+\frac{\dot\phi_L^2}2\Big) \equiv\frac{\rho_\phi}{M_P^2},
    \end{eqnarray}
enters the stage. This equation can be rewritten in the manifestly Friedmann form with the effective Planck mass $M_{\rm eff}(\rho)$ depending on the full matter density $\rho$,
    \begin{eqnarray}
    &&\frac{\dot a_L^2}{a_L^2}+\frac1{a_L^2}=
    \frac{\rho}{3M^2_{\rm eff}(\rho)}, \\
    &&M^2_{\rm eff}(\rho)=
    \frac{M_P^2}2\left(\,1+\sqrt{1
    -\frac{\beta\,\rho}{12\pi^2M_P^4}}\,\right),  \label{effPlanck0}
    \end{eqnarray}
and $\rho$ together with the inflaton energy density $\rho_\phi$ includes the primordial radiation of the CFT cosmology \cite{bigboost}
    \begin{eqnarray}
    &&\rho=\rho_\phi+\frac{R}{a^4_L}, \nonumber \\
    &&R=3M_P^2\Big(\,C-\frac{B}2\,\Big)=\frac1{2\pi^2}
    \sum\limits_\omega\frac\omega{e^{\eta\omega}\mp1}.
    \end{eqnarray}

The further evolution for large $t$ consists in the fast quasi-exponential expansion during which the primordial radiation gets diluted, the inflaton field decays by a conventional exit scenario and goes over into the quanta of {\em conformally non-invariant} fields produced from the vacuum. They get thermalized and reheated to give a new post-inflationary radiation. Thus the primordial radiation of numerous conformal species and the inflaton energy  get replaced with the radiation of non-conformal particles, $\rho\to\rho_{\rm rad}$, which give rise to the radiation dominated Universe. With $\rho_{\rm rad}$ dropping down below the sub-Planckian energy scale, $\rho_{\rm rad}\ll M_P^4/\beta$, the effective Planck mass (\ref{effPlanck0}) tends to its Planckian value, $M_{\rm eff}(\rho)\to M_P$, and one obtains a standard general relativistic inflationary scenario for which the initial conditions were prepared by our CFT garland instanton \cite{bigboost,DGP/CFT}.

The parameters of this inflation scenario are determined by the properties of the inflaton potential at the nucleation point $\phi_*=\phi(\tau_*)$. For the quadratic potential (\ref{quadpot}) at $\tau_*$ we have $\phi_*=\phi_0-\Delta_\phi$ and $V_*\equiv V(\phi_*)=3M_P^2H^2$ in view of $\dot\phi_*=0$. Therefore the expressions for inflationary slow roll parameters at this point read
    \begin{eqnarray}
    &&\eta_*\equiv M_P^2\frac{V''_*}{V_*}
    =-\frac{\mu^2}{3H^2},                      \label{eta}\\
    &&\epsilon_*\equiv\frac{M_P^2}2
    \left(\frac{V'_*}{V_*}\right)^2=
    \frac12\left(\frac{\Delta_\phi}{M_P}\right)^2
    \left(\frac{\mu^2}{3H^2}\right)^2.
    \end{eqnarray}
Note that they are related by the equation
    \begin{eqnarray}
    \epsilon_*=
    \frac12\left(\frac{\Delta_\phi}{M_P}\right)^2
    \eta_*^2,                                      \label{epsvseta}
    \end{eqnarray}
which implies for $\Delta_\phi\sim M_P$ the typical relation $\eta_*\sim\epsilon_*^2$ for slow roll parameters in the Starobinsky model \cite{Starobinskymodel} or in the model with a non-minimally coupled inflaton \cite{nonminimal}.

With $H$ given by (\ref{Hmn}) the second slow roll smallness parameter (\ref{eta}),
    \begin{eqnarray}
    &&|\,\eta_*|=\frac23\,B\mu^2
    +\frac43\frac{n^2}{m^2},                  \label{etamn}
    \end{eqnarray}
fails to be small unless $B\mu^2\ll 1$ and $m^2\gg n^2$. This gives additional ground for the scale factor oscillations to be much faster than those of the inflaton, $m\gg n$, and we consider this limit below.

\subsection{Fast oscillations of the scale factor}
For $m\gg n\geq 1$ the number $m$ of the scale factor oscillations during the full period means that we consider $m$-fold instanton garland for which the energy scale is very close to the upper bound of the range (\ref{Hbound}) \cite{slih}
    \begin{eqnarray}
    H^2\simeq\frac1{2B}
    \left(1-\frac{\ln^2 m^2}{2\pi^2m^2}\right). \label{Hcrit}
    \end{eqnarray}
The Euclidean solutions of the CFT cosmology in this limit can be called slow-roll ones, because the rate of change of the scale factor is much higher than that of the inflaton field. The Euclidean version of the slow-roll regime is, however, rather peculiar, because in contrast to the Lorentzian case with monotonically changing variables here the scale factor and inflaton are oscillating functions of time. The details of these solutions including, in particular, the derivation of this asymptotics for $H^2$ (first given in \cite{slih}) are presented in Appendix B. Here we give a simplified overview of these solutions and their relation to the conventional slow roll parameters of inflation in Lorentzian theory.

Comparison of (\ref{Hcrit}) with (\ref{Hmn}) implies that
    \begin{eqnarray}
    n^2\simeq\frac{B\mu^2}{4\pi^2}\,\ln^2m^2.
    \end{eqnarray}
Since $n\geq 1$ the lower bound on $m$ is exponentially high, and the ratio $n/m$ in (\ref{etamn}) becomes exponentially small,
    \begin{eqnarray}
    &&m\geq e^{\pi/\sqrt{B\mu^2}},     \label{mbound}\\
    &&\frac{n}m\simeq
    \frac{\sqrt{B\mu^2}}{2\pi}\,
    \frac1m\,\ln m^2
    \leq e^{-\pi/\sqrt{B\mu^2}}.
    \end{eqnarray}
As a result the solution becomes very close to the upper quantum gravity scale -- the cusp of the curvilinear triangle on Fig.1 and the corresponding slow roll smallness parameter (\ref{etamn}) expresses in terms of the quantity $B\mu^2=B|\,V_*''|$,
    \begin{eqnarray}
    H^2\simeq\frac1{2B},\quad
    \eta_*\simeq-\frac23\,B\mu^2.   \label{parameters}
    \end{eqnarray}
In view of the known CMB data for $n_s=1-6\epsilon_*+2\eta_*\simeq 1+2\eta_*\simeq 0.96$, this quantity is thus supposed to be also very small, $B\mu^2\sim 0.01$. Now, bearing in mind that $\Delta\leq\varepsilon$ with
    \begin{eqnarray}
    \varepsilon\simeq\frac{\ln^2 m^2}{2\pi^2m^2}
    \leq\frac2{B\mu^2}e^{-2\pi/\sqrt{B\mu^2}},
    \end{eqnarray}
we have the admissible range of $\Delta$, $\Delta\leq \frac2{B\mu^2}\,e^{-2\pi/\sqrt{B\mu^2}}$. This range is, however, further reduced by the requirement of the harmonic oscillator approximation (\ref{approximation0}), $\Delta\ll\varepsilon$, and finally reads
    \begin{eqnarray}
    \Delta\ll
    \frac2{B\mu^2}e^{-2\pi/\sqrt{B\mu^2}}. \label{Deltabound}
    \end{eqnarray}
This bound is stronger than the requirement of a valid harmonic oscillator approximation (\ref{bound1}), $\mu\gg\omega\Delta$, or $\Delta^2\ll \varepsilon\frac{\mu^2}{4H^2}\simeq \varepsilon \frac{B\mu^2}2$, which reads  as $\Delta\ll e^{-\frac\pi{\sqrt{B\mu^2}}}$, because in the full range of values of $B\mu^2$ the last bound is higher than (\ref{Deltabound}) by the factor of $\frac{B\mu^2}2\,e^{\pi/\sqrt{B\mu^2}}\geq\frac{\pi^2e^2}2\gg1$.

This establishes the range of the amplitude of scale factor oscillations $\Delta$. To match with the slow roll smallness parameter $\eta_*\sim-0.01$ it should be exponentially small
    \begin{eqnarray}
    \Delta\ll
    \frac4{3|\eta_*|}
    e^{-2\sqrt2\pi/\sqrt{3|\eta_*|}}. \label{Deltabound1}
    \end{eqnarray}
Below we will estimate the bound on the first smallness parameter $\epsilon_*$ in the CFT cosmology modelling initial conditions for  the non-minimal Higgs inflation. It requires the knowledge of the amplitude of the inflaton field oscillations $\Delta_\phi$. As we will see,  this amplitude will have a typical sub-Planckian value, so that a typical relation $\epsilon_*\sim\eta_*^2$ characteristic of the Starobinsky model or the model of the non-minimal Higgs inflation will hold and signify that $\epsilon_*$ adds a negligible contribution to the CMB spectral parameter and provides a very small tensor to scalar ratio $r=16\epsilon_*$.

\section{Non-minimal Higgs inflation: the mechanism of hill like inflaton potential}
In what follows we want to advocate that the CFT cosmology can serve as a source of initial conditions for the non-minimally coupled Higgs inflation model \cite{BezShap,RGH} which, together with the Starobinsky model of $R^2$-inflation \cite{Starobinskymodel}, is considered as one of the most promising models fitting the CMB data \cite{WMAP,Planck}. There is a twofold reason for that because, firstly, the Higgs inflation model at the quantum level has a natural mechanism of forming a hill-top potential and, secondly, it provides a relation $\varepsilon_*\sim\eta_*^2\ll|\eta_*|$, $\eta_*<0$, which establishes a strong link between the observable value of the CMB spectral parameter
    \begin{equation}
    n_s=1-6\epsilon_*+2\eta_*\simeq 1+2\eta_*\simeq 0.96
    \end{equation}
and the value of the Higgs mass discovered at LHC \cite{we,BezShap1,Wil,BezShap3}. This relation, as we will see below, will be provided by the bound on the amplitude of the inflaton oscillations $\Delta_\phi$ in the underbarrier regime.

The inflationary model with a non-minimally coupled Higgs-inflaton $H$, $\varphi^2=H^\dag H$, as any other semiclassical model, has a  low-derivative part of its effective action (appropriate for the inflationary slow-roll scenario),
    \begin{eqnarray}
    &&\varGamma_{\rm Higgs}[\,g_{\mu\nu},\varphi\,]=\int d^{4}x\,g^{1/2}
    \Big(V(\varphi)-U(\varphi)\,R(g_{\mu\nu})\nonumber\\
    &&\qquad\qquad\qquad\quad+
    \frac12\,G(\varphi)\,
    (\nabla\varphi)^2\Big)\ .   \label{effaction1}
    \end{eqnarray}
Its coefficient functions contain together with their tree-level part the logarithmic loop corrections and include the dependence on the UV normalization scale $\mu$,
    \begin{eqnarray}
    &&V(\varphi)=\frac\lambda{4}\varphi^4+
    \frac{\lambda \mbox{\boldmath$A$}}{128\pi^2}\,
    \varphi^4
    \ln\frac{\varphi^2}{\mu^2},             \label{effpot}\\
    &&U(\varphi)=
    \frac{M_P^2}2+\frac{\xi\varphi^{2}}2+
    \frac{\mbox{\boldmath$C$}}{32\pi^2}\varphi^2
    \ln\frac{\varphi^2}{\mu^2},           \label{effPlanck}\\
    &&G(\varphi)=1+\frac{\mbox{\boldmath$F$}}{32\pi^2}
    \ln\frac{\varphi^2}{\mu^2}.           \label{phirenorm}
    \end{eqnarray}
Numerical coefficients $\mbox{\boldmath$A$},\mbox{\boldmath$C$},\mbox{\boldmath$F$}$ are determined by contributions of quantum loops of all particles and represent beta functions of the corresponding running coupling constants -- quartic self-coupling $\lambda$, non-minimal coupling of the Higgs field to curvature $\xi$ -- and anomalous dimension of $\varphi$. In the one-loop approximation these logarithmic corrections comprise the Coleman-Weinberg potential for the Higgs inflaton and the relevant correction to the non-minimal curvature coupling. When $\xi\gg 1$ the inflationary stage and the corresponding CMB parameters critically depend only on $\mbox{\boldmath$A$}$ and the part of $\mbox{\boldmath$C$}$ linear in $\xi$, $\mbox{\boldmath$C$}=3\xi\lambda$ \cite{we,RGH}. Other coefficients, the normalization scale $\mu$ inclusive, are irrelevant in the leading order of the slow roll expansion.

Inflation and its CMB are easy to analyze in the Einstein frame of fields $\hat g_{\mu\nu}$,
$\phi$, in terms of which the action (\ref{effaction1})
        \begin{eqnarray}
        &&\varGamma_{\rm Higgs}[\,g_{\mu\nu},\varphi\,]
        =\hat\varGamma_{\rm Higgs}[\,\hat g_{\mu\nu},\phi\,]\nonumber \\
        &&\equiv\int d^{4}x\,\hat g^{1/2}
        \left(\hat V(\phi)-\frac{M_P^2}2\,R(\hat g_{\mu\nu})+
        \frac12\,(\hat\nabla\phi)^2\right)
           \label{Eframe1}
        \end{eqnarray}
has a minimal coupling of the inflaton to curvature, $\hat U=M_P^2/2$, a canonical normalization of the inflaton field, $\hat G=1$, and a new inflaton potential,
        \begin{eqnarray}
        \hat{V}(\phi)=\left.\left(\frac{M_P^2}{2}\right)^2
        \frac{V(\varphi)}{U^2(\varphi)}
        \,\right|_{\,\varphi=
        \varphi(\hat\varphi)}~.               \label{hatV}
        \end{eqnarray}
These Einstein frame fields are related to the Jordan frame of Eq.(\ref{effaction1}) by the equations
        \begin{eqnarray}
        \hat g_{\mu\nu}=\frac{2U(\varphi)}
        {M_P^2}g_{\mu\nu},\,\,\,\,
        \left(\frac{d\phi}{d\varphi}
        \right)^2
        =\frac{M_P^2}{2}\frac{GU+3U'^2}{U^2}.   \label{Eframe}
        \end{eqnarray}

Due to the presence of leading logarithmic terms the Einstein frame potential starts decreasing to zero for large values of $\varphi$,
        \begin{eqnarray}
        \hat{V}(\phi)\simeq\frac{\pi^2M_P^4}{9\lambda}
        \frac{\mbox{\boldmath$A$}}{\xi^2\ln(\varphi/\mu)}\to 0,\quad \varphi\to\infty.
        \end{eqnarray}
In this limit the original Jordan frame field $\varphi$ and the Einstein frame field $\phi$ are related by $\varphi\simeq \frac{M_P}{\sqrt\xi}e^{\phi/\sqrt{6}M_P}$, so that this asymptotics looks like $\hat{V}(\phi)\simeq \frac{\pi^2\sqrt{6}}{9\lambda\xi^2\phi}\mbox{\boldmath$A$} M_P^5\to 0$, $\phi\to\infty$. Of course, this behavior cannot be extrapolated to infinity, because semiclassical expansion fails at transplanckian  energies, but the maximum of the potential, which is reached at some $\bar\phi$, $\hat{V}'(\bar\phi)=0$, $\bar\varphi^2\simeq\mu^2\exp\left(16\pi^2/3\lambda\right)$,
corresponds for $\xi\gg 1$ to a small subplanckian value of energy
        \begin{eqnarray}
        \hat{V}(\bar\phi)\simeq\frac{\mbox{\boldmath$A$}}{96\,\xi^2}\,M_P^4
        \ll M_P^4.
        \end{eqnarray}
Therefore the potential starts bending down in the domain where the semiclassical expansion is still applicable and where it acquires a shape of suitable for our hill-top inflation scenario.

In fact a similar qualitative behavior holds in any order of loop expansion, because in the leading logarithm approximation of $l$-th loop order both $V(\varphi)$ and $U(\varphi)$ grow like $l$-th power of the logarithm while their ratio $V/U^2$ in the Einstein frame potential (\ref{hatV}) decreases like $(\ln(\varphi/\mu))^{-l}$. This property was confirmed numerically within RG resummation of leading logarithms in \cite{RGH} in the model of non-minimal Higgs inflation, which is illustrated by the plot of $\hat V$ on Fig.3.

\begin{figure}[h]
\centerline{\epsfxsize 9cm \epsfbox{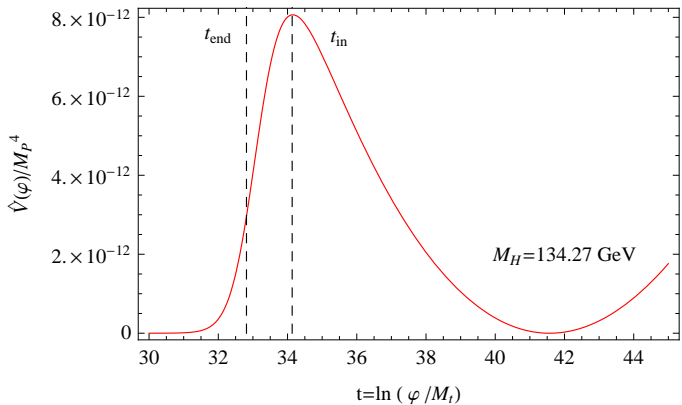}}
\caption{\small Hill-like shape of the renormalization group improved effective potential in the non-minimal Higgs inflation model as a function of the logarithmic scale with the Higgs field $\varphi$ in units of the top quark mass $M_t$ \cite{RGH}. Higgs mass value $M_H=134.27$ GeV compatible with the CMB data at the one-loop order \cite{RGH} at the two-loop order matches with the currently observed at LHC value $M_H\simeq 126$ GeV \cite{BezShap3}. The inflation domain is marked by dashed lines. \label{Fig.3}}
\end{figure}

The logic of application of our hill-top scenario to the non-minimal Higgs inflation model implies a set of careful transitions between the original Jordan frame and the Einstein frame on the FRW background. We start with the full action containing the Einstein-Hilbert part, the non-minimal Standard model part and the large  $\mathbb{N}$ CFT part
    \begin{eqnarray}
    S[\,g_{\mu\nu},H,...\varPhi\,]&=&
    S_{EH+SM}[\,g_{\mu\nu},H,...\,]\nonumber \\
    &&+
    S_{CFT}[\,g_{\mu\nu},\varPhi\,],
    \end{eqnarray}
where
    \begin{eqnarray}
    &&S_{EH+SM}[\,g_{\mu\nu},H,...\,]=
    \int d^{4}x\,g^{1/2}
    \left(\frac{\lambda\varphi^4}4\right.\nonumber \\
    &&\qquad\qquad
    \left.-\frac{M_P^2+\xi\varphi^2}2\,R+
    \frac12\,(\nabla\varphi)^2+...\right)
    \end{eqnarray}
includes the contributions of the Higgs field $\varphi^2\equiv H^\dag H$ non-minimally coupled to the metric and of the other Standard model fields denoted by ellipses. Quantization of this theory, which we perform in the original {\em Jordan} frame\footnote{Quantization of scale invariant theories in the Einstein frame with the asymptotically shift invariant inflaton potential, allegedly, stabilizes radiative corrections \cite{BezShapSib}. Flatness of the potential, however, results not only in the smallness of vertices, but also makes the theory effectively massless. This undermines the gradient expansion which is a corner stone of the Higgs inflation model and brings to life strong infrared effects \cite{IR}. This and the other properties of shift-invariant potentials in the Einstein frame make us to prefer quantization in the Jordan frame.}, results in the effective action of the gravitating Higgs model (\ref{effaction1}) and the effective action of the CFT sector (\ref{GammaCFT})
    \begin{eqnarray}
    &&S_{EH+SM}[\,g_{\mu\nu},H,...\,]\to \varGamma_{\rm Higgs}[\,g_{\mu\nu},\varphi\,],\\
    &&S_{CFT}[\,g_{\mu\nu},\varPhi\,]\to\varGamma_{CFT}[\,g_{\mu\nu}].
    \end{eqnarray}

Now we rewrite the Higgs effective action in the Einstein frame of fields $\hat g_{\mu\nu}$ and $\phi$ according to (\ref{Eframe1}) and take it on the FRW background, $\hat\varGamma_{\rm Higgs}[\,\hat g_{\mu\nu},\phi\,]|_{FRW}=\hat\varGamma_{\rm Higgs}[\,\hat a,\hat N,\phi\,]$. Then it reads as a classical part of the action (\ref{minimal_inflaton}) but in terms of the Einstein frame scale factor $\hat a$ and lapse $\hat N$ and with quantum corrected potential $\hat V(\phi)$ of the above hill-like shape,
    \begin{eqnarray}
    &&\hat\varGamma_{\rm Higgs}[\,\hat a,\hat N,\phi\,]=m_P^2\oint d\tau\,\hat N\,\left\{\vphantom{\frac11}-\hat a \hat a'^2
    -\hat a\right.\nonumber \\
    &&\qquad\qquad\qquad\quad\left.+\frac{\hat a^3 }{3M_P^2}\,
    \left(\hat V(\phi)+\frac12\,
    \phi'^2\right)\,\right\},            \label{Gamma1}\\
    &&\hat a'=\frac1{\hat N} \frac{d\hat a}{d\tau},\,\phi'=\frac1{\hat N}\frac{d\phi}{d\tau}.
    \end{eqnarray}

The CFT quantum effective action which was obtained on the FRW background in the original {\em Jordan} frame by the conformal transformation method, $\varGamma_{CFT}[\,g_{\mu\nu}]|_{FRW}=\varGamma_{\rm CFT}[\,a,N\,]$, has the following form in terms of the original $a$ and $N$
    \begin{eqnarray}
    &&\varGamma_{\rm CFT}[\,a,N\,]=
    Bm_P^2\oint d\tau N\left(\frac{a'^2}{a}
    -\frac{a'^4}{6 a}+\frac1{2a}\right)\nonumber \\
    &&\qquad\qquad\qquad+ F(\eta),\quad
    a'=\frac1N\frac{da}{d\tau}.   \label{Gamma2}
    \end{eqnarray}
Now we have to rewrite this CFT action in terms of the Einstein frame variables, $\varGamma_{\rm CFT}[\,a,N\,]=\hat\varGamma_{\rm CFT}[\,\hat a,\hat N,\phi\,]$, so that the full effective action -- the sum of (\ref{Gamma1}) and (\ref{Gamma2}) will be parameterized in the Einstein frame.  Under the replacement
    \begin{eqnarray}
    a^2=\frac{M_P^2}{2U}\,\hat a^2\simeq \frac{M_P^2}{\xi\varphi^2}\,\hat a^2,\,\,
    N^2=\frac{M_P^2}{2U}\,\hat N^2\simeq \frac{M_P^2}{\xi\varphi^2}\,\hat N^2,
    \end{eqnarray}
(in which we disregard logarithmic corrections in $U$ or absorb them in the running coupling constants) the conformal time remains unchanged in view of its local conformal invariance, $\eta=\oint d\tau N/a=\oint d\tau\hat N/\hat a$, and only the derivative of the scale factor undergoes the transition
    \begin{eqnarray}
    a'=\sqrt{U}\,\frac{d}{\hat N d\tau}\frac{\hat a}{\sqrt U}\simeq
    \hat a'-\frac{\hat a}{\varphi}\frac{d\varphi}{\hat N d\tau}=
    \hat a'-\frac{\phi'}{\sqrt{6}M_P}\hat a.
    \end{eqnarray}
This can be interpreted as as an additional contribution of the conformal anomaly due to the transition from the Jordan frame to the Einstein frame.

For a slowly varying scalar field,
    \begin{eqnarray}
    \frac{|\phi'|}{\sqrt 6 M_P}
    \ll\frac{|\hat a'|}{\hat a},     \label{slowroll}
    \end{eqnarray}
this contribution is small, so that $\hat\varGamma_{\rm CFT}[\,\hat a,\hat N,\phi\,]\simeq\\
\varGamma_{\rm CFT}[\,\hat a,\hat N\,]$, and the full action $\hat\varGamma_{\rm Higgs}[\,\hat a,\hat N,\phi\,]\\+\varGamma_{\rm CFT}[\,\hat a,\hat N\,]$ takes the form of the minimally coupled inflaton action (\ref{minimal_inflaton}) in terms of hatted variables. One can apply to it the above analysis. The additional restriction $|\dot a|/a\gg|\dot\phi|/\sqrt 6 M_P$ (we omit hats from now on) implies the bound
    \begin{eqnarray}
    \frac{\Delta_\phi^2}{M_P^2}\, \frac{\mu^2}{H^2}\ll\frac{\Delta^2}\varepsilon.  \label{cond3}
    \end{eqnarray}
This is stronger than the bounds (\ref{bound20}) and (\ref{bound3}). On account of the first bound (\ref{bound1}), $\mu\gg\omega\Delta$ or $\Delta^2/\varepsilon\ll\mu^2/H^2$, it takes the form
    \begin{eqnarray}
    \frac{\Delta_\phi^2}{M_P^2}\ll 1.    \label{cond4}
    \end{eqnarray}
This suppresses $\epsilon$ in (\ref{epsvseta}) even below its value in the Starobinsky or Higgs inflation models and makes the estimate for the spectral parameter $n_s$ even less sensitive to the value of $\epsilon$.

\section{CFT driven cosmology and the Starobinsky $R^2$ model}
Non-minimal Higgs inflation model is very similar to the Starobinsky $R^2$-model \cite{Starobinskymodel} from the viewpoint of inflation theory predictions \cite{BezrukovGorbunov:HiggsvsStar}.
Here we will show that inclusion of this $R^2$-model into the full action of the CFT cosmology is very important because it not only supplies a dynamical degree of freedom with the hill like inflaton potential, but also provides the theory with a necessary finite renormalization (\ref{renormaction}) of the trace anomaly coefficient $\alpha$.

Note that the contribution of the conformal anomaly to the effective actions (\ref{effaction0}) and (\ref{minimal_inflaton}) originates from the Wess-Zumino procedure of integrating the trace anomaly (\ref{anomaly}) along the orbit of the conformal group $g_{\mu\nu}=e^\sigma \bar g_{\mu\nu}$. The resulting Wess-Zumino action for $\sigma$ is just the difference of effective actions calculated on two members of this orbit $g_{\mu\nu}$ and $\bar g_{\mu\nu}$. It reads \cite{anomalyaction}
    \begin{eqnarray}
    &&\varGamma_{CFT}[\,g\,]-\varGamma_{CFT}[\,\bar g\,]=
    \frac{1}{64\pi^2}\int d^4x \,
    \bar g^{1/2} \Big[\,\gamma\, \bar C_{\mu\nu\alpha\beta}^2\nonumber \\
    &&\quad+\beta\,\Big(\bar E-\frac{2}{3}\,\bar\Box \bar R\Big)\Big]\,\sigma            \nonumber\\
    &&\quad
    +\frac\beta{64\pi^2}\int d^4x \,\Big[\,\bar g^{1/2}\,\sigma{\cal \bar D}\sigma
    -\frac19\,g^{1/2}R^2+\frac19\,\bar{g}^{1/2}\bar R^2\Big] \nonumber\\
    &&\quad-
    \frac{\alpha}{384\pi^2}\,
    \int d^4x\,(g^{1/2}R^2-
    \bar{g}^{1/2}\bar R^2),              \label{RTF}
    \end{eqnarray}
where all barred quantities are built in terms of $\bar g_{\mu\nu}$ and ${\cal \bar D}$ is the barred version of the fourth-order Paneitz operator ${\cal D} = \Box^2 + 2R^{\mu\nu}\nabla_{\mu}\nabla_{\nu} -
    \frac{2}{3} R\,\Box + \frac{1}{3}(\nabla^{\mu}R)\,\nabla_{\mu}$.

This expression has an important property -- with $\alpha=0$ the only higher order (quartic) derivatives of $\sigma$, contained in the combination $\bar g^{1/2}\sigma\bar{\cal D}\sigma-\frac19\,g^{1/2}R^2$ in the second line above, completely cancel out, and the resulting Wess-Zumino action does not acquire extra higher-derivative degrees of freedom \cite{slih}. With a nonzero $\alpha$ the same property holds for the renormalized action $\varGamma^R_{CFT}[\,g\,]$ in (\ref{renormaction}). The increment of this action along the orbit of the local conformal group becomes $\alpha$-independent and acquires the following {\em minimal} form
    \begin{eqnarray}
    &&\varGamma_{CFT}[\,g\,]\to \varGamma^R_{CFT}[\,g\,]
    =\varGamma_{CFT}[\,g\,]\nonumber \\
    &&\qquad\qquad\qquad+\frac\alpha{384\pi^2}\int d^4x\,
    g^{1/2}\,R^2(g).                   \label{renormaction1}
    \end{eqnarray}
The increment of this action along the orbit of the local conformal group becomes $\alpha$-independent and acquires the following {\em minimal} form
    \begin{eqnarray}
    &&\varGamma^R_{CFT}[\,g\,]-\varGamma^R_{CFT}[\,\bar g\,]=
    \frac{\gamma}{4(4\pi)^2}\int d^4x \bar g^{1/2}\,\sigma
    \,\bar C_{\mu\nu\alpha\beta}^2             \nonumber\\
    &&
    +\frac{\beta}{2(4\pi)^2}\int d^4x \bar g^{1/2} \left\{\,\frac{1}{2}\,
    \sigma \bar E-\Big(\bar R^{\mu\nu}-\frac12\bar g^{\mu\nu}\bar
    R\,\Big)\,\partial_\mu\sigma\,\partial_\nu\sigma \right.  \nonumber\\
    &&\left.-
    \,\frac12\,\bar\Box\sigma\,
    (\bar\nabla^\mu\sigma\,\bar\nabla_\mu\sigma)
    -\frac18\,(\bar\nabla^\mu\sigma\,
    \bar\nabla_\mu\sigma)^2\right\},         \label{minimal}
    \end{eqnarray}
where again all barred quantities are built in terms of the metric $\bar g_{\mu\nu}$. On the FRW background spacetime of $S^1\times S^3$-topology with the metric (\ref{FRW}) and conformally related metric of the Einstein static universe $d\bar s^2\equiv ds^2/a^2=d\eta^2+d\Omega_{(3)}^2$, $d\eta=N\,d\tau/a$, this difference of effective actions -- the conformal anomaly contribution to (\ref{effaction0}) -- equals
    \begin{eqnarray}
    &&\big(\,\varGamma^R_{CFT}
    -\bar\varGamma^R_{CFT}\,\big)\,\big|_{FRW}\nonumber \\
    &&\qquad\qquad\qquad=
    B m_P^2\int_{S^1} d\tau\,N \left(\frac{a'^2}{a}
    -\frac{a'^4}{6 a}\right).
    \end{eqnarray}

Another part of the full CFT action in (\ref{effaction0}) is the effective action of CFT fields on the Einstein universe, which consists of the free energy $F(\eta)$ and the linear in $\eta$ contribution of the Casimir energy $E_0$. Prior to the finite renormalization (\ref{renormaction}) it equals
    \begin{eqnarray}
    \varGamma_{CFT}[\,\bar g\,]\,\big|_{FRW}=F(\eta)+E_0\eta,\,\, E_0=\frac38\left(\beta-\frac\alpha2\right),
    \end{eqnarray}
where a particular dependence of the Casimir energy on the coefficients of the conformal anomaly was observed in the cosmological context \cite{Casimir_energy} and universally derived in the class of conformally-flat spacetimes from the normalization of this energy to zero in flat spacetime \cite{universality}.
\footnote{Flat spacetime can also be connected to the Einstein universe by another special conformal transformation, the relevant vacuum stress tensors being related via the conformal factor of this transformation. This gives the dependence of $E_0$ on $\beta$ and $\alpha$ \cite{universality}.} Since $(\alpha/384\pi^2)\int d^4x\,\bar g^{1/2}\bar R^2=3\alpha\eta/16$, the renormalization (\ref{renormaction}) also leads to the renormalization of the Casimir energy
    \begin{eqnarray}
    E_0\to E_0^R\equiv\frac38\,\beta=\frac{Bm_P^2}2,
    \end{eqnarray}
which acquires a particular value $\sim B/2$ corresponding to $\alpha=0$. This value of $E_0^R$ contributes to the full action of the model (\ref{effaction0}), and it is essentially responsible for the particular properties of the garland instantons \cite{slih,hatch}.\footnote{This value of the Casimir energy guarantees that the limiting case of the garland instantons with the radiation constant $C\to 0$ corresponds to chains of ``touching" exact spheres $S^4$ whose contribution is suppressed to zero by infinite positive effective action \cite{hatch}. This suppression excludes the infrared catastrophe of the Hartle-Hawking no-boundary state leading, as is well known, to the anti-intuitive conclusion that the quantum origin of an infinitely large universe is infinitely more probable than that of the finite one. Other values of the Casimir energy would lead to conical singularities in this set of instantons and would leave the resolution of this issue ambiguous.}

This is, of course, equivalent to the well-known statement that the coefficient of $\Box R$ in the trace anomaly can always be renormalized to zero by the counterterm quadratic in Ricci scalar \cite{BD}, which is admissible from the viewpoint of UV renormalization due to its locality. However, we want to emphasize here that this renormalization results in twofold consequences -- CFT quantum corrections preserving the non-dynamical nature of the scale factor and a particular value of the Casimir energy $\sim Bm_P^2/2=3\beta/8$ which universally expresses via the topological (Gauss-Bonnet) coefficient in the conformal anomaly. In this respect, $\varGamma_{CFT}$ in the formalism of Sects.1-4 above should be everywhere labeled by $R$, which implies that this is the renormalized action which incorporates these properties. Below we show that this finite renormalization can be enforced by the inclusion of the Starobinsky model $R^2$-term as a part of the full action.

Indeed, the renormalization (\ref{renormaction}) can be viewed as a replacement of the Einstein-Hilbert action by the action of the Starobinsky model \cite{Starobinskymodel} with the coupling constant $\xi$ of the curvature squared term
    \begin{eqnarray}
    &&S_{EH}[\,g_{\mu\nu}]\equiv -\frac{M_P^2}2\int d^{4}x\,g^{1/2}
    \,R\; \to\; S_\xi^{\rm Star}[\,g_{\mu\nu}],\nonumber \\
    &&S^{\rm Star}_\xi[\,g_{\mu\nu},\varphi\,]
    \equiv\int d^{4}x\,g^{1/2}
    \left(-\frac{M_P^2}2\,R
    -\frac\xi4\,R^2\right).     \label{Star}
    \end{eqnarray}
This allows one to rewrite the full action of the theory with a generic set of conformal fields as a combination of the Starobinsky model with a particular value of the coupling constant $\xi=\alpha/96\pi^2$ and the renormalized CFT action of the non-dynamical dilaton $a^2=e^\sigma$,
    \begin{eqnarray}
    &&\varGamma\equiv S_{EH}+\varGamma_{CFT}= S_\xi^{\rm Star}\,\big|_{\,\xi
    =\alpha/96\pi^2}+\varGamma_{CFT}^R.
    \end{eqnarray}

In its turn the Starobinsky action, which contains an additional conformal degree of freedom, can be rewritten in terms of the scalar-tensor theory with a non-minimal curvature coupling of the auxiliary scalar field $\varphi$,
    \begin{eqnarray}
    &&S_\xi^{\rm Star}[\,g_{\mu\nu}]\;\to\;S^{\rm Star}_\xi[\,g_{\mu\nu},\varphi\,]\nonumber \\
    &&\,\,=\int d^{4}x\,g^{1/2}
    \left\{-\frac{M_P^2}2\,
    \left(1+\xi\frac{\varphi^2}{M_P^2}\right)R
    +\frac{\xi\varphi^4}4\,\right\}.
    \end{eqnarray}
On the solution of the equation of motion for $\varphi$ this scalar field expresses in terms of the scalar curvature $\varphi^2=R$ and, thus, recovers the original purely metric representation of the Starobinsky model. This brings us to the equivalent formulation of the full effective action as
    \begin{equation}
    \varGamma[\,g_{\mu\nu},\varphi\,]= S_\xi^{\rm Star}[\,g_{\mu\nu},\varphi\,]\,
    \big|_{\,\xi=\alpha/96\pi^2}
    +\varGamma_{CFT}^R[\,g_{\mu\nu}].   \label{action10}
    \end{equation}

Naively here the field $\varphi$ does not have a kinetic term, though of course it is hidden in the non-minimal coupling of $\varphi$ to gravity. Transition to the Einstein frame according to (\ref{Eframe}) with $G=0$ and $U=(M_P^2+\xi\varphi^2)/2$ gives the relation between the Jordan frame fields and those of the Einstein frame $\phi$ and $\hat g_{\mu\nu}$ and the Einstein frame potential $\hat V$
    \begin{eqnarray}
    &&1+\xi\frac{\varphi^2}{M_P^2}=
    e^{\frac{2|\,\phi\,|}{M_P\sqrt6}},\quad g_{\mu\nu}=
    e^{-\frac{2|\,\phi\,|}{M_P\sqrt6}}\,
    \hat g_{\mu\nu},                       \label{phivarphi}\\
    &&\hat V\!=\!\frac14\,\frac{\xi\varphi^4}{\big(\,1
    +\xi\varphi^2/M_P^2\,\big)^2}\!
    =\!\frac{M_P^4}{4\xi}
    \left(\,1-e^{-\frac{2|\,\phi\,|}{M_P\sqrt6}}\,\right)^2\!.  \label{hatV1}
    \end{eqnarray}
Here the modulus of $\phi$ is chosen in order to cover the range of negative $\varphi$ by negative values of $\phi$.

If the slow roll condition (\ref{slowroll}) holds, then this transition leaves the CFT part of the action unchanged similarly to the case of the Higgs inflation model discussed above. Therefore, on the FRW background the action (\ref{action10}) takes the form of our original effective action (\ref{minimal_inflaton})(where $a$ and $\phi$ are understood as hatted Einstein frame variables) with the minimal dynamical inflaton having the potential (\ref{hatV1}). This potential is asymptotically shift invariant at large $\phi$, but logarithmic radiative corrections render it a hill-top shape by the mechanism discussed in the previous section. Thus, we can apply the dynamical inflaton scenario considered above.

Moreover, the inclusion of the Starobinsky model becomes indispensable if we put forward as a guiding principle a necessity to preserve the garland instantons and their dynamical inflaton generalization. This is because its $R^2$ term is the only means to render a non-dynamical nature of the dilaton (scale factor) mode and a particular value of the Casimir energy $\sim B/2$. The restriction on the CFT model for this to hold is the positivity of the overall value of $\alpha$. It is necessary for the positivity of $\xi$ in (\ref{Star}) -- an admissible range of $\xi$ providing {\em unitarity} -- absence of ghosts in the Starobinsky model (quite paradoxically corresponding to a negative definite Euclidean $R^2$-action). This restriction is rather mild, because for low conformal spins $\alpha$ is dominated by a positive contribution of vector particles which should only be not outnumbered by scalar bosons and fermions,
    \begin{eqnarray}
    \alpha=\frac1{90}\,\big(-\mathbb{N}_0-3\mathbb{N}_{1/2}+
    18\mathbb{N}_{1}\big),
    \end{eqnarray}
where $\mathbb{N}_0$, $\mathbb{N}_{1/2}$ and $\mathbb{N}_1$ are the numbers of scalar, spinor (Weyl) and vector particles respectively.

Another restriction follows from the observation that the inflaton potential plateau in (\ref{hatV}) has over-Planckian scale $24\pi^2 M_P^4/\alpha$, because, in contrast to the Higgs model, $\xi=\alpha/96\pi^2$ is small for a typical overall value of $\alpha=O(1)$. This difficulty can be circumvented by demanding a large value of $\alpha\gg 24\pi^2$, which requires either a huge number of vector bosons or higher spin conformal particles (see below). Another possibility is to start, from the very beginning, instead of (\ref{action10}) with the model consisting of the large $\xi$ Starobinsky model and the CFT theory
    \begin{equation}
    \varGamma\equiv S_\xi^{\rm Star}+\varGamma_{CFT}= S_{\xi+\alpha/96\pi^2}^{\rm Star}+\varGamma_{CFT}^R, \label{double_role}
    \end{equation}
so that even with the negative total $\alpha$ the sub-Planckian  bound on $\hat V$,
    \begin{eqnarray}
    \hat V
    =M_P^4
    \frac{(\,1-e^{-2|\phi|/M_P\sqrt6}\,)^2}{4\xi
    +\frac{\alpha}{24\pi^2}},                          \label{hatV2}
    \end{eqnarray}
would imply the bound on $\xi$, $4\xi+\alpha/24\pi^2\gg 1$. Then Eq.(\ref{double_role}) can be interpreted as the way $R^2$-gravity plays its double role -- part of it performs finite renormalization of $\alpha$ to zero, $\varGamma_{CFT}\to\varGamma_{CFT}^R$, while the rest of it generates a dynamical inflaton feeding the CFT scenario with the potential (\ref{hatV2}).

Needless to say that a similar mechanism can be attained by using a pure $R^2$ term in Eqs.(\ref{Star})-(\ref{action10}) without the Einstein-Hilbert term. This would correspond to a classically scale-invariant theory in which the Einstein-Hilbert term arises by dimensional transmutation due to quantum corrections -- UV renormalizable models of this type (with additional $R_{\mu\nu}^2$ terms in the action) were recently considered in \cite{Strumia,Jones}.

\section{Hierarchy problem and higher spin conformal fields}
The major difficulty in the construction of a realistic inflationary model via SLIH scenario is the hierarchy problem. Its inflation scale $H^2\simeq 1/2B$ requires the parameter $B$ to be exceedingly large in order to belong to the sub-Planckian domain compatible with the CMB data. When expressed in terms of the coefficient $\beta$ of the overall trace anomaly (\ref{anomaly}) the corresponding inflaton potential read
    \begin{eqnarray}
    V_*=3M_P^2H^2\sim \frac{3M_P^2}{2B}
    =\frac{12\pi^2}\beta\, M^4_P.
    \end{eqnarray}
For the Higgs inflation with a large non-minimal coupling $\xi\sim 10^4$ the (Einstein frame) energy density at the start of inflation is $V_*\sim 10^{-11}M_P^4$ \cite{we}, so that for the total beta we must have
    \begin{eqnarray}
    \beta\sim 10^{13}.     \label{beta_bound}
    \end{eqnarray}
In order to reach this value with the conventional low spin particle phenomenology characteristic of the Standard Model one would need unrealistically high numbers of conformal invariant scalar bosons $\mathbb{N}_0$, Dirac fermions $\mathbb{N}_{1/2}$ and vector bosons $\mathbb{N}_{1}$ in the expression for $\beta$
    \begin{eqnarray}
    \beta=\frac1{180}\,\big(\mathbb{N}_0+11 \mathbb{N}_{1/2}+
    62 \mathbb{N}_{1}\big).                \label{100}
    \end{eqnarray}

Hidden sector of so numerous low spin and very weakly interacting particles does not seem to be realistic. However, unification of interactions inspired by the ideas of string theory, holographic duality \cite{AdS/CFT0} and higher spin gauge theory \cite{Vasiliev} suggests that this hidden sector might contain conformal higher spin (CHS) fields \cite{GKPST,Tseytlin,GiombiKlebanovTseytlin}, so that the total value of $\beta$ consists of the additive sum of their partial contributions
    \begin{eqnarray}
    \beta=\sum_s\beta_s \mathbb{N}_s.                \label{101}
    \end{eqnarray}
Recently there was essential progress in the theory of these fields. In particular, it was advocated in \cite{GKPST,Tseytlin} that the values of $\beta_s$ can be explicitly calculated for conformal fields of an arbitrary spin $s$, described by totally symmetric tensors and (Dirac) spin-tensors,
    \begin{eqnarray}
    &&\beta_s=\frac1{360}\,\nu_s^2(3+14\nu_s),
    \quad \nu_s=s(s+1),\nonumber\\ 
    &&s=1,2,3,...\,,\label{boson}\\
    &&\beta_s=\frac1{720}\,
    \nu_s(12+45\nu_s+14\nu_s^2),\nonumber \\
    &&\nu_s=-2\Big(s+\frac12\Big)^2,
    s=\frac12,\frac32,\frac52,...\,,   \label{fermion}
    \end{eqnarray}
where $\nu_s$ is their respective number of dynamical degrees of freedom -- polarizations (negative for fermions). Though these fields serve now basically as a playground for holographic AdS/CFT duality issues and suffer from the problems of perturbative unitarity, which is anticipated to be restored only at the non-perturbative level,\footnote{Lack of perturbative unitarity is associated with the fact that CHS fields are not free from ghosts -- their kinetic operator contains higher derivatives of order $2s$. Manifestation of this fact is the possibility of negative energies, negative values of $\beta$ like in the case of conformal fermions starting with spin $s=3/2$, etc.} it is worth trying to exploit them as a possible solution of hierarchy problem in the CFT driven cosmology.

Strong motivation for this is that partial contributions of individual higher spins rapidly grow with the spin as $s^6$, so that the tower of spins up to some large $S$ generates the total value of $\beta$
    \begin{eqnarray}
    \beta\simeq \frac7{180}\int_0^S ds\,s^6=\frac{S^7}{180}  \label{beta}
    \end{eqnarray}
(for simplicity we consider only bosons and assume that every higher spin species is taken only once, $\mathbb{N}_s=1$). Therefore, in order to provide the hierarchy bound (\ref{beta_bound}) the maximal spin should be $S\sim 100$, which corresponds to the following estimate of the total number of particle modes (polarizations) in the hidden sector of the theory
    \begin{eqnarray}
    \nu=\sum_s\nu_s\simeq\int_0^S ds\,s^2\sim 10^6. \label{nu}
    \end{eqnarray}

\section{Conclusions}
Thus, for a wide range of parameters satisfying the bounds (\ref{approximation0}) and (\ref{bound4}) the CFT cosmology with a dynamical inflaton having a hill-like potential has instantons which are very close to the garland type solutions of the model with a primordial cosmological constant. These garland instantons are described by the approximation of two coupled oscillators and generate a new type of hill-top inflation depicted on Fig.2.

Exponentially high number of the garland instanton folds $m$, corresponding to the upper bound on the effective cosmological constant in the range (\ref{Hbound})-(\ref{lowerCbound}), guarantees smallness of slow roll parameters $\epsilon$ and $\eta$ beginning with their values $\epsilon_*$ and $\eta_*$ at the Euclidean-Lorentzian transition point. They turn out to satisfy the relations $\epsilon\sim\eta^2$ and $\eta<0$, characteristic of the well-known non-minimal Higgs inflation or Starobinsky $R^2$ gravity models. Hill-like potential in these models can be generated by logarithmic loop corrections to their tree-level asymptotically shift-invariant potential, so that with the bounds (\ref{cond3}) or (\ref{cond4}) on the amplitude and frequency of oscillations of cosmological instantons, CFT cosmology can be regarded as a source of initial conditions for the Higgs inflation with a strong non-minimal coupling or the Starobinsky $R^2$ gravity. The bound (\ref{cond4}) implies the relation $\epsilon\ll\eta^2$ at the onset of inflation, which even stronger bounds the tensor to scalar ratio in the CMB signal of these models.

It was shown that the $R^2$-gravity becomes indispensable if we put forward as a guiding principle a necessity to preserve special properties of garland instantons in the CFT scenario. This is because the $R^2$ term is the only means to render a non-dynamical -- and therefore stable non-ghost -- nature of the scale factor mode and a particular value of the Casimir energy in (\ref{effaction0}) and (\ref{bootstrap}). In fact it plays a double role -- a part of it serves as a renormalization tool for the conformal anomaly effective action, while the rest of it contributes a dynamical inflaton feeding the CFT scenario with the asymptotically shift-invariant potential which is converted by radiative corrections to the hill-like shape.

A major difficulty with this scenario is the hierarchy problem between the subplanckian scale of inflation and the CFT scale $M_P^4/\beta$. Possible solution consists in invoking a hidden sector of conformal higher spin fields which can generate a large value of trace anomaly coefficient $\beta$ (see Eq.(\ref{beta_bound})). Though this mechanism of a large $\beta$ is very vulnerable to criticism regarding unitarity problem, problem of naturalness and fine tuning, it fits presently very popular ideas of string theory, higher spin gauge theory \cite{Vasiliev} and holographic duality \cite{GKPST}, and it will be considered in much detail in the coming publication \cite{CHScosmology}.

For the inverted quadratic potential (\ref{quadpot}) garland instantons establish a strong correlation between a small magnitude of the inflationary parameter $|\eta|\simeq 0.02$ and exponentially high bound (\ref{mbound}) on the number of garland folds, $m\geq \exp(\pi/\sqrt{B\mu^2})\simeq \exp(\pi\sqrt2/\sqrt{3|\eta|})$. This is another serious difficulty of the CFT model, which makes the admissible range of instantons exceedingly narrow (\ref{Deltabound1}) and too close to the upper bound of the cosmological constant range (\ref{Hbound}). In particular, this makes thermal effects of primordial CFT radiation absolutely negligible, because the thermal contribution to the CMB red tilt $\Delta n_s^{\rm thermal}$ derived in \cite{CMBA-theorem} turns out to be {\em exponentially} suppressed by an {\em exponentially} high value of $m$. In the $l$'th CMB multipole it approximately reads \cite{CMBA-theorem}
    \begin{eqnarray}
    \Delta n_s^{\rm thermal}(l)\simeq
    -\frac{20\,l\,m^{1/3}}{(3\,\tilde\beta)^{1/6}}
    \exp\left[-
    \frac{10\,l}{(3\,\tilde\beta)^{1/6}}
    \,m^{1/3}\right],                     \label{Deltans}
    \end{eqnarray}
where $\tilde\beta$ is the average value of $\beta$ per one effective conformal degree of freedom $\tilde\beta\simeq\sum_s \beta_s \mathbb{N}_s/\sum_s \nu_s \mathbb{N}_s$. Smallness of this contribution for large $m$ would unfortunately destroy experimental verifiability of this model, because the thermal pattern of the power spectrum -- temperature of the CMB temperature \cite{CMBA-theorem} -- seems to be the only observable effect which distinguishes density matrix intial conditions of the Universe from its initial pure quantum state.

Fortunately, this problem can be solved by considering potentials more realistic than (\ref{quadpot}) with their convexity $|V''_*|$ at the onset of inflation point being smaller than the one at their top $\mu^2=|V''_0|$. For such potentials correlation between small $|\eta|$ and large $m$ no longer holds, and the estimate for $B\mu^2$ can be raised to yield a moderate $O(1)$ lower bound on $m$. This makes the thermal tilt (\ref{Deltans}) largely determined by the value of $\tilde\beta$ following from the CFT particle content of the model.

Remarkably, for the tower of CHS fields up to $S=100$, chosen above as a solution of the hierarchy problem, the estimate for the average value of $\beta$ per one conformal degree of freedom amounts in view of (\ref{beta_bound}) and (\ref{nu}) to the value $\tilde\beta\sim 10^7$  (we assume that each higher spin species enters only once, $\mathbb{N}_s=1$). With this value of $\tilde\beta$ and $m=O(1)$ the thermal correction (\ref{Deltans}) to the spectral index appears in its third decimal order, $\Delta n_s^{\rm thermal}\sim -0.001$, which is within the reach of a coming technology. This means that a potential resolution of the hierarchy problem via CHS fields simultaneously makes measurable a thermal contribution to the CMB red tilt, which is complementary to the conventional tilt of the CMB spectrum \cite{ChibisovMukhanov}.

The final remark regarding the CMB signal of the CFT model is that its precise power spectrum has not yet been found, except the thermal contribution (\ref{Deltans}) found in \cite{CMBA-theorem}. Conventional dependence of CMB parameters on $\epsilon$ and $\eta$ \cite{ChibisovMukhanov} might be modified by a nontrivial speed of sound caused by nontrivial kinetic terms in the effective action (\ref{minimal}) -- the issue subject to current research \cite{CFTCMB}.

\section*{Acknowledgements}
The authors are grateful to F.Bezrukov, M.Shaposhnikov, S.Sibiryakov and A.Starobinsky for fruitful and thought-provoking correspondence and discussions. They also greatly benefitted from helpful discussions with J.B.Har\-tle, D.Blas, C.Burgess, C. Deffayet, J.Garriga, D.Gorbu\-nov, A.Panin, P.Stamp, N.Tsamis, A.Tseytlin, W.Unruh, \\A.Vilenkin and R.Woodard. A.B. is grateful for hospitality of Theory Division, CERN, where this work was initiated, and Pacific Institute for Theoretical Physics, UBC, where this work was accomplished.  His work was also supported by the RFBR grant No. 14-01-00489 and by the Tomsk State University Competitiveness Improvement Program. The work of A.K. and D.N. was supported by the RFBR grant No.14-02-01173.

\appendix
\renewcommand{\thesection}{Appendix \Alph{section}.}
\renewcommand{\theequation}{\Alph{section}.\arabic{equation}}

\section{(An)harmonic scale factor oscillations}

The parameters of the solution $\varepsilon$ and $\Delta=\sqrt{1-4CH^2}$ in the inflaton model become slowly varying functions of time in the vicinity of the upper boundary of the triangular domain of Fig.1, $\Delta\to 0$. However, the rate of change of $\Delta$ is much faster than that of $\varepsilon$, $|\dot\varepsilon|\ll|\dot\Delta|$, because
    \begin{eqnarray}
    &&\dot\varepsilon=-2(1-\varepsilon)W\frac{\dot a}a,\\
    &&\dot\Delta=-\frac{1-\Delta^2}{\Delta}W\frac{\dot a}a\simeq-\frac1\Delta W\frac{\dot a}a,\quad \Delta\ll 1,
    \end{eqnarray}
where we used Eq.(\ref{dotH}) which implies
    \begin{eqnarray}
    &&\frac{\dot H}{H}=W\frac{\dot a}a,\\
    &&W\equiv\frac{\dot\phi^2}{2H^2M_P^2}.
    \end{eqnarray}
Slow rate of change of $W$ during the period of oscillation of $a$, $\tau_-\leq\tau\leq\tau_+$, $|\dot W|\ll\omega|W|$, allows us to integrate the above equation for $\Delta$ and obtain in view of (\ref{a_vs_y})
    \begin{eqnarray}
    &&\Delta^2(\tau)=\Delta^2(\tau_+)-W\ln\frac{a^2(\tau)}{a^2_+(\tau_+)}\nonumber \\
    &&\simeq\Delta^2(\tau_+)+Wy\Delta(\tau)+W\Delta(\tau_+)+O(W\Delta^2).
    \end{eqnarray}
Note that the quantities (\ref{apm}), $a_\pm^2\equiv (1\pm\Delta)/2H^2$,  via $H=H(\tau)$ are now functions of time $a_\pm=a_\pm(\tau)$, and $a_+(\tau_+)$ is a real constant -- the maximum value of the scale factor during a given period of oscillations. The above quadratic equation can be solved for the quantity
    \begin{eqnarray}
    \lambda\equiv\frac\Delta{W}
    \end{eqnarray}
in terms of the variable $y$ running between the values $\pm1$ during the individual period of the scale factor oscillations
    \begin{eqnarray}
    \lambda(y)=\frac{y}2+\sqrt{\frac{y^2}4+\lambda_++\lambda_+^2},
    \quad \lambda_+\equiv\frac{\Delta(\tau_+)}
    {W(\tau_+)}.                            \label{lambdaofy}
    \end{eqnarray}
From this relation it follows that during this period of oscillation $\lambda$ (and correspondingly $\Delta$) change from the minimal value $\lambda_+$ to the maximal value $\lambda_++1$,
    \begin{eqnarray}
    \lambda_+\leq\lambda\leq\lambda_++1,\quad
    \Delta_+\leq\Delta\leq\Delta_++W.
    \end{eqnarray}
A fast change of $\Delta$ means that the differentiated version of the equation (\ref{a_vs_y}), $\frac{a^2}B=\frac{1+\Delta y}{1-\varepsilon}$, takes the form
    \begin{eqnarray}
    &&\frac{a\dot a}B=\frac1{1-\varepsilon}\,
    \left(-\frac12\,\Delta\dot y+\frac{W}{2\Delta}\,y\,\frac{\dot a}a\right),
    \end{eqnarray}
where we disregard $\dot\varepsilon$ terms. Solving this equation with respect to $\dot a/a$ we have
    \begin{eqnarray}
    &&\frac{\dot a}a=-\frac12\,\frac{\Delta \dot y}{1-\frac{W}{2\Delta}
    y}+O(\Delta^2/W).
    \end{eqnarray}
If we use this relation in Eq.(\ref{dota}) whose right hand side in terms of $y$ reads as
    \begin{eqnarray}
    \dot a^2=\frac{\Delta^2}{2\varepsilon}(1-y^2)\,
    \Big(1+O(\Delta/\varepsilon)\Big),
    \end{eqnarray}
then
    \begin{eqnarray}
    &&\dot y^2=\omega^2\left(1-\frac{y}{2\lambda(y)}\right)^2
    \left(1-y^2\right)\,\Big(1+O(\Delta/\varepsilon)\Big),\\
    &&\omega=\sqrt{\frac{2(1-\varepsilon)}{\varepsilon B}}=\frac{2H}{\sqrt{1-2BH^2}}=\frac{2H}{\sqrt\varepsilon},
    \end{eqnarray}
where $\omega$ is the same slowly varying in time frequency as (\ref{omega}). This equation is the generalization of Eq. (\ref{doty}) by the factor $(1-y/2\lambda(y))^2\neq 1$ taking into account rapid variations of $\Delta$. This factor which renders the oscillations of $y$ very anharmonic does not, however, change the period of these oscillations
    \begin{eqnarray}
    T\!&=&\!2\int\limits_{-1}^{1}\frac{dy}{\dot y}\!
    \simeq\!
    \frac2\omega\int\limits_{-1}^{1}\!\frac{dy}{\sqrt{1-y^2}}\!
    \left(1+\frac{y}{\sqrt{y^2+4\lambda_++4\lambda_+^2}}\right)\nonumber \\
    &=&
    \frac{2\pi}\omega.
    \end{eqnarray}
In the leading order approximation in $\Delta/\varepsilon\ll1$ this period coincides with the period of harmonic oscillations of (\ref{y}) for any positive value of $\lambda_+$. In fact, with $\lambda(y)$ given by (\ref{lambdaofy}) the equation for $y(\tau)$,
    \begin{eqnarray}
    \frac{\dot y}{\sqrt{1-y^2}} \left(1+\frac{y}{\sqrt{y^2+4\lambda_+
    +4\lambda_+^2}}\right)=\omega,
    \end{eqnarray}
can be easily integrated to give
    \begin{eqnarray}
    y&=&\frac{\frac1{1+2\lambda_+}
    -\cos(\omega\tau)}
    {\sqrt{1+\frac1{(1+2\lambda_+)^2}
    -\frac2{1+2\lambda_+}\cos(\omega\tau)}}\nonumber \\
    &=&
    \frac{(1+2\lambda_+)\sin^2\frac{\omega\tau}2-\lambda_+}
    {\sqrt{\lambda_+^2+(1+2\lambda_+)\sin^2\frac{\omega\tau}2}}.
    \end{eqnarray}
In the limit $\lambda_+\to\infty$ this solution reduces to the harmonic law (\ref{y}), whereas for $\lambda_+\to 0$ (that is $\Delta\ll W$) it reads as the function not differentiable at $\sin\frac{\omega\tau}2=0$,
    \begin{eqnarray}
    y=\left|\,\sin\frac{\omega\tau}2\right|.
    \end{eqnarray}
The range of $\tau$ where $-1\leq y\leq 0$ in this limit shrinks to these singular points with $\sin\frac{\omega\tau}2=0$.
\\
\\
\section{``Slow roll" Euclidean solutions in CFT cosmology}
``Slow roll'' solutions in the Euclidean regime are analogous to the Lorentzian solutions, except that the scale factor and the inflaton are oscillating instead of respectively monotonically growing and decreasing, and the scale factor oscillations are much faster than those of the inflaton field.
Oscillations of $\phi$ with a constant frequency $\mu$ and rapid oscillations of $a$ with a slowly varying frequency $\omega$ should be commensurable. Then the integer number $n$ of full oscillations of $\phi$ within the time period  $T=2\pi n/\mu$ corresponds to $m$, $m\gg n$, oscillations of $a$. The resulting total increase of the phase of these oscillations by $2\pi m$ is given by the integral
    \begin{eqnarray}
    \int\limits_0^{2\pi n/\mu} d\tau\,\omega(\tau)=2\pi m.
    \end{eqnarray}
For a constant $\omega$ this relation implies the commensurability equality (\ref{commensurable}), but with a slowly varying $\omega(\tau)$ small corrections arise. By iterating the solution of the exact equation
    \begin{eqnarray}
    H^2(\tau)=H^2_*+\int\limits_{\tau_*}^{\tau} d\tau'\,\frac{\dot\phi^2}{M_P^2}\,\frac{\dot a}a(\tau'),
    \end{eqnarray}
where $H^2_*=H^2(\tau_*)$ is the value of $H$ at the beginning of the full period $\tau_*=0$ -- the turning point of the solution, one finds leading corrections in $\Delta$ and $\Delta_\phi$ to time dependent $H^2$ and $\varepsilon=1-2BH^2$,
    \begin{eqnarray}
    &&H^2(\tau)\simeq H^2_*+\frac{\omega_*\Delta_*}4
    \frac{\Delta_\phi^2}{M_P^2}\,
    \mu^2\left(
    \frac{\sin^2(\mu-\frac{\omega_*}2)\tau}{2\mu-\omega_*}\right.\nonumber\\
    &&\qquad\qquad\left.-\frac{\sin^2(\mu+\frac{\omega_*}2)\tau}{2\mu+\omega_*}
    +\frac{\sin^2\frac{\omega_*\tau}2}{\omega_*}
    \right).
    \end{eqnarray}
Then one can estimate the total period of the instanton in units of the conformal time. This period is the sum of $m$ contributions of individual periods of $a$-oscillations. Each of them equals $\pi\sqrt{2\varepsilon_i}$, where $\varepsilon_i$ is taken at $\tau_i$ belonging to the $i$-th individual period of duration $\Delta\tau_i=T_i$ which according to Appendix A is equal to $2\pi/\omega_i$ with $\omega_i=2H_i/\sqrt{1-2BH^2_i}$,
    \begin{eqnarray}
    &&\eta\simeq\pi\sum\limits_{i=1}^m\sqrt{2\varepsilon_i}=
    \pi\sum\limits_{i=1}^m\sqrt{2\varepsilon_i}
    \frac{\Delta\tau_i}{T_i}\nonumber\\
    &&\quad\simeq
    \frac12\int_0^{2\pi n/\mu} d\tau\,\omega\sqrt{2\varepsilon}=\sqrt2\int_0^{2\pi n/\mu} d\tau\,H(\tau).
    \end{eqnarray}
As a result the leading corrections to the frequency commensurability equation (\ref{commensurable}) and $\eta$ have the following form
    \begin{eqnarray}
    &&\mu m\simeq\omega_*n\left(1-\frac{\Delta_*}{2\varepsilon_*}
    \frac{\Delta_\phi^2}{M_P^2}
    \frac{\mu^2}{H_*^2}\frac{n^2}{m^2}\right),\\
    &&\eta\simeq\pi m\sqrt{2\varepsilon_*}
    \left(1-\frac{\Delta_*}2
    \frac{\Delta_\phi^2}{M_P^2}
    \frac{\mu^2}{H_*^2}\frac{n^2}{m^2}\right),
    \end{eqnarray}
where all starred quantities are taken at $\tau_*$. Suppression of these corrections by the factor $\Delta\Delta_\phi^2/M_P^2$ is obvious from the dynamical equation (\ref{dotH}) for $H^2$, while the extra suppressing factor $n^2/m^2\ll 1$ comes from integration of slow $\phi$-harmonics modulated by rapid oscillations of the scale factor. As we will see below, these corrections of the magnitude
    \begin{eqnarray}
    \frac\Delta\varepsilon\times
    \frac{n^2}{m^2}                \label{leadingcorr}
    \end{eqnarray}
will be negligibly small, so that they can be discarded in the leading order, and moreover the star labels of $H$, $\varepsilon$ and $\Delta$ can also be omitted in this approximation. Therefore, as in Sect.3.1 we have
    \begin{eqnarray}
    \mu m\simeq\omega n, \quad \eta\simeq\pi m\sqrt{2\varepsilon}.            \label{B7}
    \end{eqnarray}

From the definition of the variable $\Delta$ and the bootstrap equation (\ref{bootstrap}) we have
    \begin{eqnarray}
    &&H^2=\frac{1-\Delta^2}{4C}=
    \frac1{2B}\frac{1-\Delta^2}{1+R},   \label{B8}\\
    &&R\equiv\frac{2F'}{Bm_P^2},
    \end{eqnarray}
where $R$ is the ratio of the thermal energy $F'/m_P^2$ to the Casimir energy $B/2$. Therefore
    \begin{eqnarray}
    \varepsilon=\frac{R+\Delta^2}{1+R},
    \end{eqnarray}
and the admissible range of parameters $\Delta\leq\varepsilon$ implies that $\Delta(1+R)\leq R+\Delta^2$ or
    \begin{eqnarray}
    \Delta\leq R.      \label{DeltaleqR}
    \end{eqnarray}

From Eqs.(\ref{B7}) up to corrections $O(\Delta_\phi^2/M_P^2\times\Delta\times n^2/m^2)$ we obtain the closed equation for $\eta$,
    \begin{eqnarray}
    \frac{\eta^2}{2\pi^2m^2}
    \simeq\frac{R(\eta)+\Delta^2}{1+R(\eta)}, \label{Beta}
    \end{eqnarray}
and the expression for $H^2$
    \begin{eqnarray}
    &&H^2=
    \frac1{2B}\,\frac1{1+2n^2/B\mu^2m^2}.
    \end{eqnarray}
Upon comparison with (\ref{B8}) this expression gives the answer for the ratio
    \begin{eqnarray}
    \frac{n^2}{m^2}\simeq\frac{B\mu^2}2 \frac{R+\Delta^2}{1-\Delta^2}.
    \end{eqnarray}
Since $n\geq 1$, this equation imposes the lower bound on $m$,
    \begin{eqnarray}
    m^2\geq\frac2{B\mu^2}\,\frac{1-\Delta^2}{R+\Delta^2},
    \end{eqnarray}
which in its turn can be rewritten as a bound on $\Delta$
    \begin{eqnarray}
    R^2\geq\Delta^2\geq\frac{\frac2{B\mu^2m^2}-R}{\frac2{B\mu^2m^2}+1},
    \end{eqnarray}
where the first inequality follows from (\ref{DeltaleqR}). One can show that the case of $R\leq 2/B\mu^2m^2$ leads to the contradiction, because the resulting value of $R$ becomes too high, $R\geq1-2\Delta^2$, and violates the low temperature limit $R(\eta)\to 0$ at $\eta\to\infty$. On the contrary, the case of
    \begin{eqnarray}
    R\geq\frac2{B\mu^2m^2}   \label{Rbound}
    \end{eqnarray}
does not lead to new bounds except the positivity of $\Delta$.

The further analyzes relies on the low temperature limit of the free energy $F(\eta)$ for large $\eta$, $F(\eta)\simeq -\mathbb{N}_0e^{-\omega_0\eta}$, where $\mathbb{N}_0$ is the number of lowest spin conformal fields having the lowest value $\omega_0$ of the energy of field-theoretical oscillator on $S^3$ ($\omega_0=1$ for a spin zero conformal field). In this limit the function $R(\eta)$ becomes
    \begin{eqnarray}
    R\simeq\frac{8\omega_0}{3\bar\eta_0}e^{-\omega_0\eta},
    \end{eqnarray}
and the solution of Eq.(\ref{Beta}) with small $R$ and $\Delta$ and a large $m$ reads $\omega_0\eta\simeq\ln m^2$, so that
    \begin{eqnarray}
    R\simeq\frac{\ln^2 m^2}{2\pi^2m^2},
    \end{eqnarray}
which indeed tends to zero for large $m\to\infty$. Then the above lower bound on $R$ gives the exponentially high lower bound on $m$,
    \begin{eqnarray}
    m\geq e^{\pi/\sqrt{B\mu^2}},
    \end{eqnarray}
and the exponentially low upper bound on $\Delta$, $R$, $\varepsilon$ and the ratio $n/m$
    \begin{eqnarray}
    &&\Delta\leq R\simeq\varepsilon
    \leq \frac2{B\mu^2}\,
    e^{-2\pi/\sqrt{B\mu^2}},    \label{RDeltabound}\\
    &&\frac{n}{m}\leq
    e^{-\pi/\sqrt{B\mu^2}}.
    \end{eqnarray}

The above bounds are consistent with the admissible range (\ref{bound4}) of $\mu$ vs $\omega$, $\omega\gg\mu\gg\omega\Delta$, which can be rewritten as
    \begin{eqnarray}
    1\gg\frac{\mu^2}{\omega^2}\simeq\frac{n^2}{m^2}\gg\Delta^2,
    \end{eqnarray}
or (in view of $n^2/m^2\simeq RB\mu^2/2$)
    \begin{eqnarray}
    \frac2{B\mu^2}\gg R\gg \frac{2\Delta^2}{B\mu^2}.
    \end{eqnarray}
This is because the first inequality coincides with the bound (\ref{RDeltabound}) for $e^{-2\pi/\sqrt{B\mu^2}}\ll1$, whereas the second inequality holds because
    \begin{eqnarray}
    &&\frac{2\Delta^2}{B\mu^2}\leq\frac\Delta{4\pi^4}
    \left(\frac{2\pi}{\sqrt{B\mu^2}}\right)^4
    e^{-2\pi/\sqrt{B\mu^2}}\nonumber \\
    &&\qquad\qquad\qquad<\frac\Delta{16\pi^4}\ll
    \Delta\leq R.
    \end{eqnarray}

Finally, these bounds justify the omission of leading corrections (\ref{leadingcorr}), because in view of $\varepsilon\simeq R$ we have
    \begin{eqnarray}
    \frac\Delta\varepsilon\,
    \frac{n^2}{m^2}\simeq\frac{R}\varepsilon\,\frac{B\mu^2}2\Delta\leq
    e^{-2\pi/\sqrt{B\mu^2}}\ll 1.
    \end{eqnarray}

\end{document}